\documentstyle[preprint,aps]{revtex}
\tighten
\draft
\begin{document}
\widetext
\input epsf
\preprint{CLNS 97/1458, HUTP-96/A056, NUB 3150}
\bigskip
\bigskip
\title{A Classification of $3$-Family Grand Unification in String Theory \\
II. The $SU(5)$ and $SU(6)$ Models}
\medskip
\author{Zurab Kakushadze$^1$\footnote{E-mail: 
zurab@string.harvard.edu} and S.-H. Henry Tye$^2$\footnote{E-mail: 
tye@hepth.cornell.edu}}
\bigskip
\address{$^1$Lyman Laboratory of Physics, Harvard University, Cambridge, MA 02138\\
and\\
Department of Physics, Northeastern University, Boston, MA 02115\\
$^2$Newman Laboratory of Nuclear Studies, Cornell University,
Ithaca, NY 14853-5001}
\date{January 14, 1997}
\bigskip
\medskip
\maketitle

\begin{abstract}
{}Requiring that supersymmetric $SU(5)$ and $SU(6)$ grand
unifications in the heterotic string theory must have $3$ chiral families,
adjoint (or higher representation) Higgs fields in the grand unified 
gauge group, and a non-abelian hidden sector, we construct 
such string models within the framework of free conformal field theory 
and asymmetric orbifolds. Within this framework,
we construct all such string models via ${\bf Z}_6$ asymmetric
orbifolds that include a ${\bf Z}_3$ outer-automorphism,
the latter yielding a level-3 current algebra for the grand unification
gauge group $SU(5)$ or $SU(6)$.
We then classify all such ${\bf Z}_6$
asymmetric orbifolds that result in models with a non-abelian hidden sector.
All models classified in this paper have only one adjoint (but no other 
higher representation) Higgs field in the grand unified gauge group.
This Higgs field is neutral under all other gauge symmetries.
The list of hidden sectors for $3$-family $SU(6)$ string models are 
$SU(2)$, $SU(3)$ and $SU(2) \otimes SU(2)$. In addition to these,
$3$-family $SU(5)$ string models can also have an $SU(4)$ hidden sector. 
Some of the models have an apparent anomalous $U(1)$ gauge symmetry.

\end{abstract}
\narrowtext

\section{Introduction}

{}For superstring theory to describe nature, it must contain the
standard model of strong and electroweak interactions as part of its low
energy ({\em i.e.}, below the string scale) effective field theory.
Among the various possibilities, grand unified theory (GUT) is particularly 
attractive because it is a truly unified theory with one gauge coupling in 
the low energy effective theory. Various GUTs in string theory have been 
extensively studied \cite{two}.
Since nature seems to have only three chiral families of quarks and
leptons, it is reasonable to impose this condition in model-building.
In earlier papers \cite{three,kt,five,class}, we have considered explicit
realization of such $3$-family GUTs in the heterotic string theory.
The number of such possibilities turns out to be quite limited.
In Ref \cite{class}, we present a classification of $3$-family $SO(10)$ and
$E_6$ grand unified heterotic string models, within the framework
of perturbative string theory.
In this paper, we present a classification of $3$-family $SU(5)$ and
$SU(6)$ grand unified heterotic string models, within a somewhat more 
restricted framework. Together with Ref \cite{class}, this paper completes our 
classification of all $3$-family grand unification within the framework of 
asymmetric orbifolds in conformal (free) field theory \cite{orb}, as 
described below.

{}For the grand unified gauge symmetry to be broken spontaneously to the 
standard model in the low energy effective field theory, it is 
well-known that an adjoint (or other appropriate representation) Higgs field 
is needed. So the requirement of such a massless Higgs field will be 
imposed in string model-building.
It is most natural to have space-time supersymmetry in string models,
which will also be imposed. This supersymmetry is expected to be broken
via dynamical supersymmetry breaking in the hidden sector, which 
requires an asymptotically-free hidden sector. This is then
transmitted to the observable sector via gravity or other
messenger/intermediate sector \cite{gaugino}.
So we are lead to impose the following
constraints on grand unified string model-building:\\
({\em i}) $N=1$ space-time supersymmetry;\\
({\em ii}) Three chiral families of fermions in the GUT gauge group;\\
({\em iii}) Adjoint (and/or other appropriate) Higgs fields in GUT;\\
({\em iv}) Non-abelian hidden sector. 

{}Imposing these constraints on string model-building within the framework 
of asymmetric orbifolds, we argue in Ref \cite{class} that, in the 
classification of $3$-family $SO(10)$ and
$E_6$ string models, that we need only 
consider ${\bf Z}_6$ asymmetric orbifolds that include a ${\bf Z}_3$ 
outer-automorphism. So the classification problem is
reduced to the classification of such ${\bf Z}_6$ asymmetric 
orbifold models that have the above properties.
This is carried out in Ref \cite{class} and in this paper. 
It is important to point out a key difference between the
construction of the $SO(10)$ and $E_6$ versus the $SU(5)$
string models. As explained in Ref \cite{class}, these GUT gauge symmetries 
are to be realized at level-3 current (Kac-Moody) algebras.
Most GUT realizations of these higher-level gauge symmetries $G_k$ employ 
the so-called diagonal embeddings 
$G_k \subset G \otimes G \otimes ... \otimes G$ $ = G^k$
(here, the superscript $k$ indicates the power, while 
the subscript $k$ indicates the level). 
A ${\bf Z}_k$ outer-automorphism of $G^k$ yields $G_k$ (plus an 
appropriate coset).
It was pointed out by Dienes and March-Russell \cite{dien} that,
for $G_k=SO(10)_3$, $(E_6)_3$, or $SU(6)_3$, this diagonal embedding 
is the maximal embedding.
So a classification of ${\bf Z}_6$ asymmetric orbifolds that yield
$3$-family $SO(10)$, $SU(6)$ and $E_6$ string models may be considered as a
classification of $3$-family $SO(10)$, $SU(6)$ and $E_6$ string models.
However, for $SU(5)_3$, there is a more economical embedding, namely, 
$SU(5)_3 \subset SU(10)_1$. This non-diagonal embedding opens the 
possibility that the ${\bf Z}_3$ outer-automorphism, and hence the 
${\bf Z}_6$ asymmetric orbifolds,
may not exhaust all possible $3$-family $SU(5)$ string models.
In this sense, the list of $3$-family $SU(5)$ string models given in 
this work may be incomplete; that is, the classification of $3$-family 
$SU(5)$ string models given in this paper includes only those that are 
reachable via the diagonal embedding.

{}In Ref \cite{class}, we gave a list of $3$-family $SO(10)$ and $E_6$ string models. Since the adjoint Higgs field is a modulus in the string moduli 
space, some $3$-family $SU(5)$ and $SU(6)$ string models can be easily 
obtained by giving the adjoint Higgs field in the $E_6$ model an 
appropriate vacuum expectation value (vev). 
In a similar fashion, $3$-family $SU(5)$ string models can be easily 
obtained from the $SO(10)$ models. Although these models are interesting 
on their own, they are rather trivial from the model-building point 
of view. In this work, we shall focus mainly on the $3$-family $SU(5)$ and 
$SU(6)$ string models that cannot be reached in this way. 
In particular, we are interested in models with an enhanced hidden sector.
Since the construction of $SU(6)$ string models is closely related to that 
of the $SU(5)$ models, their classification should go together.

{}Besides the properties that resulted from the constraints imposed in 
string model-building, all the $3$-family GUT string models share 
the following properties:\\ 
$\bullet$ there is only one adjoint (plus some lower, but no higher, 
representation) Higgs field; \\
$\bullet$ there is an intermediate/horizontal gauge symmetry.

{}In addition, there are three key phenomenologically interesting features 
that distinguish the $3$-family $SU(5)$ and $SU(6)$ string models from 
the $3$-family $SO(10)$ and $E_6$ string models: 

{}({\em i}) Recall that the $3$-family $SO(10)$ and $E_6$ string models have
either $4$ left-handed and $1$ right-handed families, or $5$ left-handed 
and $2$ right-handed families. In contrast, some of the $3$-family $SU(5)$ 
and $SU(6)$ string models have truly $3$ chiral families. For example, 
one model 
has $3$ left-handed ${\bf 10}$s, $12$ left-handed and $9$ right-handed 
${\overline {\bf 5}}$s of $SU(5)$, yielding $3$ chiral families and 
$9$ Higgs multiplets.

{}({\em ii}) In contrast to the $3$-family $SO(10)$ and $E_6$ string models,
some of the $3$-family $SU(5)$ and $SU(6)$ string models have an apparent
anomalous $U(1)$ gauge symmetry; that is, the anomaly contributions of the
$U(1)$ charged massless fermions do not add up to zero. Here, the anomaly
cancellation is accomplished by the presence of some massless scalar fields
in the model. This is the so-called Green-Schwarz mechanism in string theory, 
and this anomalous $U(1)$ gauge symmetry is expected to be broken by the
Fayet-Iliopoulos term \cite{gs}. The $U(1)$ breaking happens
at some scale slightly below the string scale, providing a natural
scale rather close to the grand unification scale naively expected.
Presumably, a number of scalar fields will develop vacuum expectation
values without breaking supersymmetry.

{}({\em iii}) A simple robust way to stabilize the dilaton expectation
value is to have a semi-simple hidden sector, with more than one gaugino
condensate \cite{kras}. So it is interesting to ask if one can get a
hidden sector with more than one gaugino condensate in 
the $3$-family GUT string model construction.
In Ref \cite{class}, we saw that the biggest (and only) asymptotically-free 
(at the string scale) hidden sector gauge symmetry was $SU(2)$.
On the other hand, there are $3$-family $SU(5)$ and $SU(6)$ string models
that  allow a hidden sector with an asymptotically-free 
$SU(2) \otimes SU(2)$ semi-simple gauge group at the string scale.
In fact, these two $SU(2)$s have different matter contents, so 
the runnings of their couplings will be somewhat different. 
This is believed to be a desirable feature. Naively, because of the large 
matter contents, these $SU(2)$s are expected to get strong at very low 
energy scales. However, this is not expected to happen. Instead,
the anomalous $U(1)$ gauge symmetry breaking is expected to give masses to
some of the $SU(2)$ multiplets. Furthermore, these $SU(2)$ 
couplings $\alpha_2$ are three times that of the grand unified gauge coupling.
So, the gauge couplings of these two $SU(2)$s will become
large at rather high scales. However, they will
become strong at different scales, each with its own gaugino
condensate \cite{gaugino}. A careful analysis of specific models 
is clearly needed to see if supersymmetry breaking happens in any of the 
models in a way that will 
stabilize the dilaton expectation value to a reasonable value\cite{kras}.

{}The list of gauge symmetries of the $3$-family $SU(6)$ string models is:\\
$SU(2)_1\otimes SU(2)_1 \otimes SU(6)_3 \otimes U(1)^3$,\\
$SU(3)_1 \otimes SU(6)_3 \otimes U(1)^3$,\\
$SU(2)_1 \otimes SU(6)_3 \otimes U(1)^4$. \\
The list of gauge symmetries of the $3$-family $SU(5)$ string models 
includes the above list where the $SU(6)_3$ is replaced by 
$SU(5)_3 \otimes U(1)$. In addition, there is a $SU(5)$ string model
with gauge symmetry, \\
$SU(4)_1 \otimes SU(2)_1\otimes SU(5)_3 \otimes U(1)^4$, \\
where the $SU(4)$ plays the role of a hidden sector.
The lists of $3$-family GUT string models are given 
in this paper and in Ref\cite{class}. Some of these models have been 
presented earlier \cite{three,kt,five}.
The massless spectra of all the 
$3$-family $E_6$ and $SO(10)$ string models are given in tables in 
Ref\cite{class}.
The massless spectra of some of the $3$-family $SU(5)$ and $SU(6)$ string 
models are given in tables in this paper. Given the orbifolds, the massless 
spectra of the remaining models can be easily written down, following 
the rules described below.

{}The rest of this paper is organized as follows.
In section II we discuss the general grounds underlying the
classification of $3$-family $SU(5)$ and $SU(6)$ string models.
Next, we turn to the construction of the models.
Since $3$-family $E_6$ and $SO(10)$ models have been constructed and
classified in Ref\cite{class} already, it is easiest if we simply 
convert the $E_6$ and $SO(10)$ models to the $SU(5)$ and $SU(6)$ models 
with appropriate Wilson lines. This approach is discussed in Section III.
The classification is given in section IV, with the massless 
spectra of the more interesting string models given in the Tables. 
In section V, we briefly discuss the moduli space of
the models and the ways they are connected to each other.
We conclude with some remarks in section VI.

\section{Preliminaries}

\bigskip

{}The general grounds on which the classification is based have been 
explained in Ref\cite{class}. First, let us briefly review them. 
Then we shall discuss the issues particular to the $SU(5)$ case.

{}({\em i}) Phenomenologically, the gauge coupling at the grand unification
scale is weak.
Since this scale is quite close to the string scale, the coupling
at the string scale is also expected to be weak. This means 
conformal field theory is valid.
So string model-building can be restricted to the four-dimensional 
heterotic string theory using conformal field theories and current
(or Kac-Moody) algebras. Our framework throughout will be
the asymmetric orbifold construction \cite{kt,orb}.

{}({\em ii}) By the three chiral families constraint,
we mean that the net number of chiral families (defined as the number of
left-handed families minus the number of the right-handed families) must be
three.

{}({\em iii}) As we mentioned earlier, the hidden sector is required so that
dynamical supersymmetry breaking may occur.
This means the hidden sector must contain a non-abelian gauge symmetry. 

{}({\em iv}) The requirement of adjoint Higgs in $N=1$ supersymmetric GUT
implies the grand unified gauge symmetry must be realized via
a higher-level current algebra.

{}An exhaustive search in the level-$2$ algebras \cite{two} 
indicates that the three chiral families plus the adjoint Higgs constraints 
require level-$3$ or higher 
current algebras for the grand unified gauge symmetry.
So it is natural to go to level-3 models. 
Since their construction typically requires a ${\bf Z}_3$
outer-automorphism, this can be part of a ${\bf Z}_3$ orbifold, which has an
odd number of fixed points. So there is a chance to construct models
with three chiral families via ${\bf Z}_3$ orbifolds.
Naively, one might expect that three-family string GUTs can be
constructed with a single ${\bf Z}_3$ twist, since the latter can,
at least in principle, be arranged to have three fixed points.
This, however, turns out not to be the case \cite{class}.
So, to obtain a model with three chiral families, we are lead to consider
asymmetric ${\bf Z}_6$ orbifolds.

{}As mentioned in section I, the common way to realize a higher-level gauge 
symmetry $G_k$ employs the so-called diagonal embeddings, that is, 
a ${\bf Z}_k$ outer-automorphism of $G^k$ yields $G_k$ (plus an
appropriate coset). It was pointed out by 
Dienes and March-Russell \cite{dien} that, for $G_k=SO(10)_3$, $(E_6)_3$, 
or $SU(6)_3$, this diagonal embedding ({\em i.e.} $G_k \subset G^k$) 
is the maximal embedding.
So a classification of ${\bf Z}_6$ asymmetric orbifolds that yield
$3$-family $SO(10)$, $SU(6)$ and $E_6$ string models may be considered as a
classification of $3$-family $SO(10)$, $SU(6)$ and $E_6$ string models.
However, this is not the case for $SU(5)_3$. Besides the diagonal embedding,
there is another embedding, namely, $SU(5)_3 \subset SU(10)_1$. Since the
central charge of $SU(10)_1$ is $9$, while that of $SU(5)^3$ is $12$, 
the embedding in $SU(10)_1$ is clearly more economical.
Since the constraint of three chiral families requires a ${\bf Z}_6$ twist,
we can ask if there is a ${\bf Z}_6$ twist of $SU(10)_1$ that yields 
$SU(5)_3$. Without even worrying about string consistency,
it is a straightforward, but somewhat tedious, exercise to show that 
no ${\bf Z}_3$ or ${\bf Z}_6$ twisting of the $SU(10)_1$ can yield $SU(5)_3$. 
This means that this non-diagonal embedding opens the possibility that the 
${\bf Z}_3$ outer-automorphism, and hence the ${\bf Z}_6$ asymmetric orbifolds,
may not exhaust all possible $3$-family $SU(5)$ string models.
In this sense, the list of $3$-family $SU(5)$ string models given in
this work may be incomplete. That is, the classification of $3$-family
$SU(5)$ string models given in this paper includes only those that are
reachable via the diagonal embedding. It is a very interesting and challenging
problem to see if $3$-family $SU(5)$ string models can be constructed via
such a non-diagonal embedding. Since $SU(5)$ is a smaller gauge group than 
the other grand unified groups, one may also want to ask the same question 
about higher-level $SU(5)$s.

{}The above discussion reduces this restricted classification problem to the
classification of ${\bf Z}_6$ asymmetric orbifold models with
the following phenomenological requirements
(that translate into stringent constraints in the actual string
model building) to be imposed: \\
$\bullet$ $N=1$ space-time supersymmetry;\\
$\bullet$ Level-3 GUT via diagonal embedding;\\
$\bullet$ Three chiral families in GUT;\\
$\bullet$ Non-abelian hidden sector.\\
It is this classification problem that we solve in Ref \cite{class} and
this paper. All of the $3$-family $SU(5)$ and $SU(6)$ string models 
(that cannot be obtained by simple spontaneous symmetry breaking of
$SO(10)$ and $E_6$ string models) are 
given in this paper, and their massless spectra are presented in the Tables.
Now the classification project consists of three steps:\\ 
({\em i}) list all suitable $N=4$ Narain models \cite{narain};\\
({\em ii}) list all ${\bf Z}_6$ asymmetric twists that can act on
each Narain model;\\
({\em iii}) work out the massless spectrum in each orbifold to check its
properties. \\
As in Ref \cite{class}, the classification only 
distinguishes different string models by their tree-level massless spectra.
It is well-known that some orbifolds, though look different at first
sight, nevertheless yield identical models. Even when their massless 
spectra are identical, some particular massless
particles may appear in different sectors in different orbifolds.
These models are simply related by $T$-duality.
It happens that their interactions are easiest to determine if we consider
both orbifolds simultaneously.

{}The gauge coupling of a given group $G$ in the model at a
scale $\mu$ below the string scale $M_s$ is related to it via:
\begin{equation}
 1/\alpha_G(\mu)= k_G/\alpha_{\mbox{string}}
 + ({b_0/{4\pi}}) \ln ({M_s^2/ \mu^2})~, \nonumber
\end{equation}
where $k_G$ is the level of the gauge group. For a $U(1)$ gauge theory,
$1/k = 2r^2$ if the $U(1)$ charge is normalized so that the lowest
allowed value is $\pm 1$ (with conformal highest weight $r^2/2$),
and $r$ is the compactification radius of the corresponding chiral world-sheet boson.
The constant $b_0$ is the one-loop coefficient of the beta-function.

{}A typical model will have some moduli that can be varied
(within some ranges) without affecting its massless spectrum.
Here we shall not distinguish such models. We will, however, discuss 
the moduli space in which these models sit.

\section{Model Construction}

\bigskip

{}Starting from a consistent heterotic string model, typically
a Narain model, a new model can be generated by performing a consistent
set of twists and shifts on it. First, these twists and shifts must be 
consistent with the symmetry of the momentum lattice.
To obtain a new consistent string model, the following conditions are
imposed on its one-loop partition function in the light-cone gauge: \\
($\em i$) one-loop modular invariance; \\
($\em ii$) world-sheet supersymmetry; and \\
($\em iii$) the physically sensible projection. \\
This last condition means the contribution
of a space-time fermionic (bosonic) degree of freedom to the partition
function counts as minus (plus) one. In all cases that can be checked, this
condition plus the one-loop modular invariance and factorization imply
multi-loop modular invariance.

{}A set of twists and shifts can be organized into a set of vectors.
The rules of consistent asymmetric orbifold model construction can then
be written as constraints on these vectors, as done in Ref \cite{kt}.
All sets of vectors given in this and the next sections,
${\em i.e.}$, the Wilson lines $(U_1, U_2)$, $A_i$ and $B_i$, 
and the ${\bf Z}_6$ orbifolds $(T_3, T_2)$, have been checked to satisfy
these consistency constraints, provided appropriate choices of the 
structure constants are picked. The structure constants dictate which sets
of states are projected out. 
The spectrum of the resulting model can be
obtained from the spectrum generating formula. It is straightforward, 
but somewhat tedious, to work out the massless spectrum in each model.
Since $3$-family $E_6$ and $SO(10)$ models have been constructed and 
classified in Ref \cite{class} already, the construction of $3$-family
$SU(5)$ and $SU(6)$ models is easiest if we simply convert the $E_6$ and 
$SO(10)$ models to them with appropriate Wilson lines.
We shall borrow heavily from Ref \cite{class}, and follow its notations.
Models labelled with $N$, $E$, $T$, $S$, and $F$ are Narain, $E_6$,
$SO(10)$, $SU(6)$ and $SU(5)$ string models respectively. 

{}There are two procedures to convert the $E_6$ and $SO(10)$ models
to $SU(5)$ and $SU(6)$ models:

($\em i$) Take one of the $E_6$ or $SO(10)$ model given in Ref\cite{class}, 
which is obtained via a particular ${\bf Z}_6$ orbifold on a specific 
Narain model. Introduce an appropriate Wilson line (compatible with 
the Narain model and the ${\bf Z}_6$ orbifold) to act on 
this Narain model to obtain a new Narain model. Such a Wilson 
line turns out to have order $3$.
Then, performing the same ${\bf Z}_6$ orbifold on this new
Narain model, a $SU(5)$ or $SU(6)$ model results. 

($\em ii$) Start from the the same original Narain model chosen above.
Instead of performing a ${\bf Z}_6$ orbifold on it, one may combine the
(order $3$) Wilson line with the ${\bf Z}_6$ orbifold to form a
${\bf Z}_6 \otimes {\bf Z}_3$ orbifold. Depending on the choice of the
new structure constants (which determine the relative phases between the 
various sectors), the action of this orbifold on the Narain model can 
yield the same $SU(5)$ or $SU(6)$ model as above.
It turns out that, sometimes, there exists an alternative choice of the 
structure constants, yielding a different $SU(5)$ or $SU(6)$ model.

{}Finally, we work out the massless spectrum of the final model to check if 
it satisfies the phenomenological constraints. By relating each $SU(5)$ or 
$SU(6)$ model to an $E_6$ or $SO(10)$ model, the tedious working out of the
massless spectrum is substantially simplified. Since the consistent
${\bf Z}_6$ orbifolds have been classified in Ref\cite{class}, the 
classification of $SU(5)$ and $SU(6)$ string models is reduced to a 
classification of the additional Wilson lines to be added. 

{}Let us illustrate these procedures by briefly describing the construction 
of the $3$-family $SU(5)$ models, namely the $F1$ and the $F2$ models. 
The first procedure involves the following steps:
({\em i}) Starting from the $N=4$ supersymmetric Narain model, 
with $SU(3) \otimes SO(8) \otimes SO(32)$ gauge symmetry, we introduce
Wilson lines to convert it to another Narain model with gauge symmetry
$SU(3) \otimes SO(8) \otimes SO(10)^3 \otimes SO(2)$, namely, 
the $N1(1,0)$ model.
({\em ii}) An appropriate ${\bf Z}_6$ orbifold on this $N1(1,0)$ model
yields the $3$-family $SO(10)$ model, with gauge symmetry 
$SU(2)_1 \otimes SU(2)_3 \otimes SO(10)_3 \otimes U(1)^3$.
This is the $T1(1,0)$ model.
({\em iii}) An appropriate Wilson line acting on the $N1(1,0)$ model yields
another Narain model, namely the $N6$ model. The same ${\bf Z}_6$ orbifold
on this $N6$ model yields the $3$-family $SU(5)$ model, namely the $F1$ model.
The rest of this section gives a more detailed description of this 
construction and the alternative construction, which yields both the
$F1$ and the $F2$ models.

{}Consider the Narain model
with the momenta of the internal bosons spanning an even self-dual
Lorentzian lattice $\Gamma^{6,22} =\Gamma^{6,6} \otimes
\Gamma^{16}$. Here $\Gamma^{16}$ is the
${\mbox{Spin}}(32)/{\bf Z}_2$ lattice. The $\Gamma^{6,6}$ is the
momentum lattice corresponding to the compactification on a six-torus
defined by $X_I= X_I +E_I$. The dot product of the vectors $E_I$
defines the constant background metric $G_{IJ}=E_I \cdot E_J$.
There is also the antisymmetric background field $B_{IJ}$.
The components of $G_{IJ}$ and $B_{IJ}$ parametrize the $36$-dimensional
moduli space of the $\Gamma^{6,6}$ lattice.
We are only interested in the subspace of the moduli space that has
(1) appropriate ${\bf Z}_3$ symmetries, upon which we have to perform a
${\bf Z}_3$ twist later, and (2) an enhanced gauge symmetry so that,
after the orbifold, the hidden sector can have maximal gauge symmetry.
With these constraints, it will be suffice to consider a two-dimensional
subspace of the general moduli space.
This subspace is parametrized by the moduli
$h$ and $f$, and the vectors $E_I$ (and also their duals ${\tilde E}^I$
defined so that $E_I \cdot {\tilde E}^J ={\delta_I}^J$) can be expressed
in terms of the $SU(3)$ root and weight vectors
${e}_i$ and ${\tilde e}^i$ ($i=1,2$):
\begin{eqnarray}
 &&E_1=(e_1,0,0),~~~E_2=(e_2,0,0),\nonumber\\
 &&E_3=(0,e_1,0),~~~E_4=(0,e_2,0),\nonumber\\
 &&E_5=(f{\tilde e}^2,-h {\tilde e}^2,ge_1),~~~E_6=(-f{\tilde e}^1,h 
{\tilde e}^1,ge_2),\nonumber
\end{eqnarray}
where $g=\sqrt{1-(f^2+h^2)/3}$.
The components of the antisymmetric tensor are chosen to be
$2B_{IJ} ={1\over 2} G_{IJ}$ for $I<J$ and $2B_{IJ} =-{1\over 2} G_{IJ}$
for $I>J$. We shall call these Narain models as $N(h,f)$.
With the above choice of the $\Gamma^{6,6}$ lattice, the Narain model 
$N(h,f)$ has the gauge symmetry $R(h,f) \otimes SO(32)$. 
Let us consider the region $0\leq h,f\leq 1$.
A generic point in the moduli space of $N(h,f)$ has $R(h,f)=
SU(3) \otimes SU(3)\otimes U(1)^2$.
There are four isolated points with enhanced gauge symmetry: \\
$\bullet$ $R(0,0)=SU(3) \otimes SU(3)\otimes SU(3)$, \\
$\bullet$ $R(1,0)=R(0,1)=SU(3) \otimes SO(8)$ and \\
$\bullet$ $R(1,1)=E_6$.

{}To be specific, let us consider the $(h,f)=(1,0)$ case.
Here we are writing the Wilson lines as shift vectors in the $\Gamma^{6,22}$
lattice. The shift vectors $U_1$ and $U_2$ to be introduced
are order-$2$ shifts that break $SO(32)$ to $SO(10)^3 \otimes SO(2)$:
\begin{eqnarray}
 &&U_1 =(e_1/2,a_1,b_1\vert\vert 0,0,0)({\bf s}\vert {\bf 0}\vert {\bf 0}\vert
 {\overline    S})~,\nonumber\\
 &&U_2 =(e_2/2,a_2,b_2 \vert\vert  0,0,0)({\bf 0}\vert {\bf s}\vert
 {\bf 0}\vert {\overline S})~.\nonumber
\end{eqnarray}
The $U_1$ and $U_2$ are order-$2$ (${\bf Z}_2$) shifts.
The first three entries correspond to the right-moving complex world-sheet
bosons.
The next three entries correspond to the left-moving complex world-sheet
bosons. Together they form the six-torus. The remaining 16 left-moving
world-sheet bosons generate the ${\mbox{Spin}}(32)/{\bf Z}_2$ lattice.
The $SO(32)$ shifts are given in the $SO(10)^3
\otimes SO(2)$ basis. In this basis, ${\bf 0}$($0$) stands for the null vector,
${\bf v}$($V$) is the vector weight, whereas
${\bf s}$($S$) and ${\overline {\bf s}}$(${\overline S}$) are the
spinor and anti-spinor weights of $SO(10)$($SO(2)$). (For $SO(2)$, $V=1$,
$S=1/2$ and ${\overline S}=-1/2$.)
Here the four component vectors $(a_1,b_1)$ and $(a_1,b_1)$ are the spinor
${\bf  s}$ and conjugate $ {\bf  c}$ weights of  $ SO(8)$, respectively.
This model has  $SU(3)\otimes SO(8)\otimes SO(10)^3 \otimes SO(2)$ gauge
symmetry. This is the $N1(1,0)$ model.

{}Next we can perform a ${\bf Z}_6$ orbifold on the above $N1(1,0)$ model.
It is easier to follow the construction by decomposing the ${\bf Z}_6$
orbifold into a ${\bf Z}_3$ twist and a ${\bf Z}_2$ twist:
\begin{eqnarray}
 &&T_3 =(\theta,\theta,\theta\vert\vert 0, \theta, \theta)
 ({\cal P} \vert  2/3)~,\nonumber\\
 &&T_2=(0, \sigma,\sigma\vert\vert  e_1/2,\sigma, \sigma)
 (0^{15} \vert 0)~. \nonumber
\end{eqnarray}
Each $\theta$ in $T_3$ is a ${\bf Z}_3$ twist (that is, a $2\pi /3$ rotation)
that acts only on a complex world-sheet boson. So the first $\theta$ acts on
the right-moving part of the $\Gamma^{2,2}$ (${\em i.e.}$, the
$SU(3)$) lattice and the corresponding oscillator excitations, while
the left-moving part is untouched. This is an asymmetric orbifold.
The $\Gamma^{4,4}$ (${\em i.e.}$, the $SO(8)$) lattice is twisted
symmetrically by the ${\bf Z}_3 \otimes {\bf Z}_3$ $\theta$ twist.
The three $SO(10)$s are permuted by the action of the
${\bf Z}_3$ outer-automorphism twist ${\cal P}$:
$\phi^I_1 \rightarrow
\phi^I_2 \rightarrow \phi^I_3 \rightarrow \phi^I_1$, where the real bosons
$\phi^I_p$, $I=1,...,5$, correspond to the $p^{\mbox{th}}$ $SO(10)$
subgroup, $p=1,2,3$. We can define new bosons $\varphi^I \equiv {1\over
\sqrt{3}}(\phi^I_1 +\phi^I_2 +\phi^I_3)$; the other ten real bosons are
complexified via linear combinations $\Phi^I \equiv {1\over
\sqrt{3}}(\phi^I_1 +\omega\phi^I_2 +\omega^2 \phi^I_3)$ and
$(\Phi^I)^\dagger \equiv {1\over \sqrt{3}}(\phi^I_1 +\omega^2\phi^I_2
+\omega \phi^I_3)$, where $\omega =\exp(2\pi i /3)$.
Under ${\cal P}$, $\varphi^I$ is invariant,
while $\Phi^I$ ($(\Phi^I)^\dagger$) are eigenstates with
eigenvalue $\omega^2$ ($\omega$).
The ${\bf Z}_3$ invariant states form irreducible representations (irreps)
of $SO(10)_3$.
Finally, string consistency requires the inclusion of the $2/3$ shift
in the $SO(2)$ lattice. 
This simply changes the radius of this world-sheet boson.
The $\sigma$ in the $T_2$ twist is a $\pi$ rotation of the
corresponding two chiral world-sheet bosons. Thus, $\sigma$ is a
${\bf Z}_2$ twist.
The left-moving momenta of $\Gamma^{2,2}$ is shifted by
$e_1/2$ (${\em i.e.}$, half a root vector), while the $\Gamma^{16}$
is left untouched.

{}The resulting $N=1$ supersymmetric $3$-family $SO(10)$ model, 
the $T1(1,0)$ model, has  
$SU(2)_1 \otimes SU(2)_3 \otimes SO(10)_3 \otimes U(1)^3$ gauge symmetry 
(the subscripts indicate the levels of the corresponding Kac-Moody algebras). 
Its massless spectrum is given in Ref\cite{class}.
This model is completely anomaly-free, and its hidden
sector is $SU(2)_1$, whereas the observable sector is $S0(10)_3 \otimes
U(1)^2$. The remaining $SU(2)_3 \otimes U(1)$
plays the role of the messenger/intermediate sector, or horizontal symmetry.
The net number of the chiral $SO(10)_3$ families in this model
is $5-2=3$. The details of the construction of this model can be found in
Ref\cite{kt}.

{}Now, the final $F1$ model can be obtained by introducing the following 
Wilson line to the $T1(1,0)$ model:
\begin{eqnarray}
  (0,0,0 \vert\vert {\tilde e}^2, -{\tilde e}^2,0)(({1\over 3}{1\over 3}
{1\over 3} {1\over 3}{2\over 3} )^3 \vert 0 )~.\nonumber
\end{eqnarray}
This Wilson line is called $A_6$ in the next section.
Here, the shift ${1\over 3}{1\over 3}{1\over 3} {1\over 3}{2\over 3}$ in
the $SO(10)$ lattice breaks the $SO(10)$ to $SU(5)^3\otimes U(1)$.
Let us introduce this Wilson line to the $N1(1,0)$ model to obtain a new
Narain model, namely, the $N6$ model. In the unshifted sector of this
$N6$ model, the gauge group is broken from $SU(3) \otimes
SO(8) \otimes SO(10)^3 \otimes SO(2)$ down to $SU(3)^2 \otimes SU(5)^3\otimes
U(1)^6$ (Note that $SO(8)$ is broken to $SU(3)\otimes U(1)^2$, and each
$SO(10)$ is broken to $SU(5)\otimes U(1)$). However, there are additional
gauge bosons that come from the shifted and inverse shifted sectors. These
are in the irreducible representations (irreps)
 $({\bf 3},{\overline{\bf 3}})(2)$ and $({\overline{\bf 3}},
{\bf 3})(-2)$ of $SU(3)\otimes SU(3)\otimes U(1)$ , where the $U(1)$ charge
(which is normalized to $1/\sqrt{6}$) is given in the parentheses. Thus, the
gauge symmetry of the $N6$ model is $SU(6) \otimes SU(5)^3 \otimes U(1)^5$.
This enhancement of gauge symmetry is made possible by breaking the $SO(10)$
subgroups, so that the resulting $F1$ model can have enhanced an
hidden sector.

{}The final $F1$ model is the ${\bf Z}_6$ orbifold $(T_3, T_2)$ of 
the $N6$ model.
When only the $T_3$ twist acts on the $N6$ model,
the resulting model has the gauge symmetry
$SU(4)_1\otimes SU(5)_3 \otimes U(1)^3$ (Note that $SU(6)$ is broken down to
$SU(4)\otimes U(1)^2$, and four of the $U(1)$s in the $N6$ models have been
removed by the $T_3$ twist). The number of chiral families of $SU(5)_3$ in
this model is $9$, as is the case for other level-3 models constructed from a
single ${\bf Z}_3$ twist \cite{class}.
Therefore, we add the $T_2$ twist to obtain a model
with three chiral families.
The $F1$ model has gauge symmetry
$SU(3)_1 \otimes SU(5)_3 \otimes U(1)^4$. Its massless spectrum is given in
the first column of Table V. They are
grouped according to where they come from, namely, the untwisted sector U,
the ${\bf Z}_3$ twisted ({\em i.e.}, $T_3$ and $2T_3$) sector T3,
the ${\bf Z}_6$ twisted ({\em i.e.}, $T_3+T_2$ and $2T_3+T_2$) sector T6,
and ${\bf Z}_2$ twisted ({\em i.e.}, $T_2$) sector T2.
Note that the $SU(3)$ subgroup in the $F1$ model 
arises as a result of the breaking $SU(4) \supset
SU(3) \otimes U(1)$. The net number of chiral families of
$SU(5)_3$ is $3$. 

{}In the alternative procedure, we combine the Wilson line $A_6$ with the
twists ($T_3, T_2$) to form the  ${\bf Z}_3 \otimes {\bf Z}_2 \otimes 
{\bf Z}_3$ orbifold generated by ($T_3, T_2, A_6$), 
acting on the $N1(1,0)$ model constructed above.
Here we find that there are additional
possibilities. Indeed, since the orders of $T_3$ and $A_6$
are the same (both of them have order three),
their respective contributions in the one-loop partition function can
have a non-trivial relative phase between them. Let this phase be $\phi(T_3,
A_6)$. This phase must satisfy  $3 \phi(T_3,A_6) =0~(\mbox{mod}~1)$
({\em i.e}, $\phi(T_3,A_6)$ can be $0,1/3,2/3$), and the
states that survive the $T_3$ projection in the $A_6$ shifted sector must
have the $T_3$ phase $\phi(T_3,A_6)$. Similarly, the states
that survive the $T_3$ projection in the inverse
shifted sector $(A_6)^{-1}$
must have the $T_3$ phase $-\phi(T_3,A_6)$. The string
consistency then requires that the states that survive the $A_6$
projection in the $T_3$ twisted sector must
have the $A_6$ phase $-\phi(T_3,A_6)$. Similarly, the states
that survive the $A_6$ projection in the inverse
twisted sector $(T_3)^{-1}$ must have the $A_6$ phase $\phi(T_3,A_6)$.
The gauge symmetry of the resulting
model depends on the choice of $\phi(T_3,A_6)$, since this phase dictates
what states in these sectors are projected out. The models with
$\phi(T_3,A_6)=0$ and $\phi(T_3,A_6)=1/3$ turn out to be equivalent, yielding
precisely the $F1$ model. The third choice $\phi(T_3,A_6)=2/3$ leads to 
a different model, which we will refer to as the $F2$ model.
The $F2$ model has gauge symmetry
$SU(2)_1\otimes SU(2)_1 \otimes SU(5)_3 \otimes U(1)^4$, with
the $SU(2)_1\otimes SU(2)_1$
forming the hidden sector. Its massless spectrum is also given in Table V. 
The $F1$ and the $F2$ models were first constructed in Ref\cite{five}. 
In contrast to the $SO(10)$ model, both of these two $SU(5)$ models
have an anomalous $U(1)$ gauge symmetry.

{}Since $A_6$ is a pure shift (without twist), the invariant sublattices 
and the numbers of fixed points in the $T_3$ and $T_2$ twists of the $F1$ 
and the $F2$ models remain the same as that in the $T1(1,0)$ model.
This allows us to borrow the construction of the $T1(1,0)$ model to help
working out the spectra of the $F1$ and the $F2$ models. 
This simplification trick will be used extensively in the next section.

\section{Classification of $SU(5)$ and $SU(6)$ Models}
\bigskip

{}In this section, we construct three-family $SU(5)$ and $SU(6)$ models 
in the framework of ${\bf Z}_6$ or ${\bf Z}_6 \otimes {\bf Z}_3$ asymmetric 
orbifolds.
To achieve this, we give a classification of ${\bf Z}_3$ Wilson lines 
that can be added to the $SO(10)$ and $E_6$ models classified in 
Ref \cite{class} to obtain the corresponding $SU(5)$ and $SU(6)$ models. 
Models labelled with $N$, $E$, $T$, $S$, and $F$ are Narain, $E_6$,
$SO(10)$, $SU(6)$ and $SU(5)$ string models respectively.
The massless spectra of many of the models are given in the tables. 
The massless spectra of the remaining models can be easliy obtained 
using the procudures presented below. 
Let us first start with the $E1$, $E2$, $T1(1,1)$ and $T2(1,1)$ models 
of Ref \cite{class}.\\
$\bullet$ The $E1$ model. The Wilson lines
\begin{eqnarray}
 &&U_1 =(0,0,0 \vert\vert e_1/2,0,0)({\bf s}\vert {\bf 0}\vert {\bf 0}\vert
 {\overline    S})~,\nonumber\\
 &&U_2 =(0,0,0 \vert\vert e_2/2,0,0)({\bf 0}\vert {\bf s}\vert {\bf 0}\vert 
{\overline S})~.\nonumber
\end{eqnarray}
acting on the N(1,1) model generates the The $N1(1,1)$ model, with
$SU(3)^2 \otimes (E_6)^3$ gauge symmetry.
Start from this $N1(1,1)$ model and perform the following twists:
\begin{eqnarray}
 &&T_3 =(\theta,\theta,\theta\vert\vert \theta,e_1/3,0)
 ({\cal P} \vert  2/3)~,\nonumber\\
 &&T_2=(\sigma, p_1,p_2\vert\vert  0,e_1/2,e_1/2)
 (0^{15} \vert 0)~.\nonumber
\end{eqnarray}
This $E1$ model has $SU(2)_1 \otimes (E_6)_3 \otimes U(1)^3$ gauge symmetry.\\
$\bullet$ The $E2$ model. This model, which is the $T$-dual of the $E1$ model,
is obtained if we perform the following twists:
\begin{eqnarray}
 &&T_3 =(0,\theta,\theta\vert\vert \theta,e_1/3,0)
 ({\cal P} \vert  2/3)~,\nonumber\\
 &&T_2=(\sigma, p_1,p_2\vert\vert  0,e_1/2,e_1/2)
 (0^{15} \vert 0)~.\nonumber
\end{eqnarray}
on the $N1(1,1)$ model. \\
$\bullet$ The $T1(1,1)$ model. The $N2(1,1)$ model is generated by the 
Wilson lines
\begin{eqnarray}
 &&U_1 =(0,e_1/2,e_1/2 \vert\vert e_1/2,0,0)({\bf s}\vert {\bf 0}\vert 
{\bf 0}\vert {\overline S})~,\nonumber\\
 &&U_2 =(0,e_2/2,e_2/2\vert\vert e_2/2,0,0)({\bf 0}\vert {\bf s}\vert 
{\bf 0}\vert {\overline S})~.\nonumber
\end{eqnarray}
acting on the $N(1,1)$ model. This $N2(1,1)$ model has
$SU(3)^2 \otimes U(1)^2\otimes SO(10)^3 \otimes SO(2)$ gauge symmetry.
Start from the $N2(1,1)$ model and perform
the same twists as in the $E1$ model. The resulting $T1(1,1)$
model has $SU(2)_1 \otimes SO(10)_3 \otimes U(1)^4$ gauge symmetry. \\
$\bullet$ The $T2(1,1)$ model. Start from the $N2(1,1)$ model and perform the
same twists as in the $E2$ model.
This model has $SU(2)_1 \otimes SO(10)_3 \otimes U(1)^4$ gauge symmetry.

{}Now, one of the following Wilson lines can be added to the above models:
\begin{eqnarray}
 &&A_1 =(0,0,0 \vert\vert 0,{\tilde e}^2,0)(({1\over 3}{1\over 3}{1\over 3}
{1\over 3}{2\over 3} )^3 \vert 0 )~,\nonumber\\
 &&A_2 =(0,0,0 \vert\vert 0,0,{\tilde e}^2)(({1\over 3}{1\over 3}{1\over 3}
{1\over 3}{2\over 3} )^3 \vert 0 )~,\nonumber\\
 &&A_3=(0,0,0 \vert\vert 0,{\tilde e}^2,-{\tilde e}^2)(({1\over 3}{1\over 3}
{1\over 3} {1\over  3}{2\over 3} )^3 \vert 0   )~.\nonumber
\end{eqnarray}  
Adding one of these Wilson lines to one of the above models yields one of 
the following $SU(6)$ and $SU(5)$ models. \\
$\bullet$ The $S2(1,0)$ model. Start from the $E1$ model and add the $A_2$ 
Wilson line. This model has $SU(2)_1 \otimes SU(2)_3 \otimes SU(6)_3 \otimes 
U(1)^3$ gauge symmetry.  The massless spectrum of the $S2(1,0)$ model is 
given in Table I. Note that the spectrum of this model is very similar to 
that of the $T2(1,0)$ model of Ref\cite{class}. In particular, the former 
can be obtained from the latter via replacing $SO(10)_3 \otimes U(1)$ 
(the last $U(1)$ normalized to $1/6$ in the second column of Table VI in 
Ref \cite{class}) by $SU(6)_3 \otimes U(1)$ (the last $U(1)$ normalized to 
$1/\sqrt{6}$ in the first column of  Table I in this paper). 
Under this substitution, ${\bf 45}(0)$ of $SO(10)_3 \otimes U(1)$ is 
replaced by ${\bf  35}(0)$ of $SU(6)_3 \otimes U(1)$,  and 
${\bf 16}(-1) + {\bf 10}(+2) +{\bf 1}(-4)$ is replaced by 
${\bf  15}(0) + {\overline {\bf 6}}(+1) +{\overline {\bf 6}}(-1)$. 
(The complex conjugates of these states are substituted similarly.)
This replacement procedure will be used extensively below.\\
$\bullet$ The $S1(1,0)$ model. Start from the $E1$ model and add the $A_1$ 
Wilson line. This model has $SU(2)_1 \otimes SU(2)_3 \otimes SU(6)_3 
\otimes U(1)^3$ gauge symmetry. The massless spectum of this model can be 
obtained from that of the $T1(1,0)$ model by the replacement procedure 
described for the $S2(1,0)$ model.\\
$\bullet$ The $S3(1,0)$ model. Start from the $E2$ model and add the 
$A_2$ Wilson line. This model has $SU(2)_1 \otimes SU(2)_3 \otimes SU(6)_3 
\otimes U(1)^3$ gauge symmetry. The massless spectum of this model can be 
obtained from that of the $T3(1,0)$ model by the replacement procedure 
described for the $S2(1,0)$ model. Note that the $S3(1,0)$ model is the 
same as the $S1(1,0)$ model, {\em i.e.}, $S3(1,0) = S1(1,0)$. \\
$\bullet$ The $S4(1,0)$ model. Start from the $E2$ model and add the 
$A_1$ Wilson line. This model has $SU(2)_1 \otimes SU(2)_3 \otimes 
SU(6)_3 \otimes U(1)^3$ gauge symmetry. The massless spectrum of this 
model can be obtained from that of the $T4(1,0)$ model by the replacement 
procedure described for the $S2(1,0)$ model. Note that $S4(1,0) = S2(1,0)$.\\
$\bullet$ The $F1(1,1)$ model. Start from the $T1(1,1)$ model and add 
either the $A_1$ or $A_2$ Wilson line (both Wilson lines give the same model). 
This model has $SU(2)_1 \otimes SU(5)_3 \otimes U(1)^5$ gauge symmetry.  
The massless spectrum of the $F1(1,1)$ model is given in Table I. 
Note that the spectrum of this model is very similar to that of the 
$T1(1,1)$ model of Ref\cite{class}. In particular, the former can be 
obtained from the latter via replacing $SO(10)_3$  by $SU(5)_3 \otimes U(1)$
(the last $U(1)$ normalized to $1/2\sqrt{15}$ in the second column of 
Table I in this paper). Under this substitution, 
${\bf 45}$ of $SO(10)_3$ is replaced by ${\bf  24}(0) +{\bf 1}(0)$ of 
$SU(5)_3 \otimes U(1)$,  ${\bf 16}$ is replaced by 
${\overline {\bf  5}}(+3) +{\bf 10}(-1) + {\bf 1}(-5)$,  ${\bf 10}$ is 
replaced by ${\bf 5}(+2)+{\overline {\bf  5}}(-2) $, and the singlet
${\bf 1}$ is replaced by ${\bf 1}(0)$. This corresponds to the 
regular breaking  $SO(10) \supset SU(5) \otimes U(1)$.\\
$\bullet$ The $F2(1,1)$ model. Start from the $T2(1,1)$ model and add 
either the $A_1$ or $A_2$ Wilson line (both Wilson lines give the same model). 
This model has $SU(2)_1 \otimes SU(5)_3 \otimes U(1)^5$ gauge symmetry.  
The massless spectrum of the $F2(1,1)$ model can be obtained from that 
of the $T2(1,1)$ model by the breaking procedure described for the 
$F1(1,1)$ model. Note that $F2(1,1) = F1(1,1)$. \\
$\bullet$ The $S1$ model. Start from the $T1(1,1)$ model and add 
the $A_3$ Wilson line. 
Choose the relative phase between the $T_3$ and $A_3$ sectors 
to be $\phi(T_3, A_3)=0$ or $1/3$ (both choices give the same model). 
This model has $SU(3)_1 \otimes SU(6)_3 \otimes U(1)^3$ gauge symmetry. 
The massless spectrum of the $S1$ model is given in Table II.\\
$\bullet$ The $S2$ model. Start from the $T1(1,1)$ model and add 
the $A_3$ Wilson line. Choose the relative phase between the $T_3$ and 
$A_3$  twisted sectors to be $\phi(T_3, A_3)=2/3$. 
This model has $SU(2)_1 \otimes SU(2)_1 \otimes SU(6)_3 \otimes U(1)^3$ 
gauge symmetry. The massless spectrum of the $S2$ model is given in Table II.\\
$\bullet$ The $S3$ model. Start from the $T2(1,1)$ model and add  
the $A_3$ Wilson line. 
Choose the relative phase between the $T_3$ and $A_3$  twisted sectors to 
be $\phi(T_3, A_3)=0$ or $1/3$ (both choices give the same model). 
This model has $SU(3)_1 \otimes SU(6)_3 \otimes U(1)^3$ gauge symmetry. 
The massless spectrum of the $S3$ model is given in Table III. 
Note that $S3 = S1$. \\
$\bullet$ The $S4$ model. Start from the $T2(1,1)$ model and add 
the $A_3$ Wilson line. Choose the relative phase between the $T_3$ and $A_3$ 
twisted sectors to be $\phi(T_3, A_3)=2/3$. This model has 
$SU(2)_1 \otimes SU(2)_1 \otimes SU(6)_3 \otimes U(1)^3$ gauge symmetry. 
The massless spectrum of the $S4$ model is given in Table III. 
Note that $S4 = S2$. \\

Here let us make a few comments for clarification. Note that adding 
the $A_3$ Wilson line to the $E1$ and $E2$ models 
does not result in three-family level-3 grand unified string models. 
This can be seen from the fact that if we add the $A_3$ Wilson line to 
the $N1(1,1)$ model of Ref\cite{class}, the resulting Narain model 
will have gauge symmetry $SU(6)\otimes SU(18)$. The corresponding 
${\bf Z}_6$ orbifold of the latter does not lead to a level-3 model.

{}The following point is important for constructing other 
$SU(6)$ models (see below). The gauge group of the $N(1,1)$ model is 
$E_6 \otimes SO(32)$. This lattice itself can be conveniently viewed 
as being constructed from the $N(0,0)$ lattice by turning on the 
following Wilson line:
\begin{eqnarray}
 U_3 =(0,0,0 \vert\vert ,{\tilde e}^2,-{\tilde e}^2, {\tilde e}^2) 
(0^{15} \vert 0 )~.\nonumber
\end{eqnarray}  
Then the corresponding $N1(1,1)$ and $N2(1,1)$ models can be viewed 
as being constructed from the $N(0,0)$ lattice by turining on the $U_3$ 
Wilson line along with the appropriate $U_1$ and $U_2$ Wilson lines. 
Now, the Wilson lines $A_1$, $A_2$, $A_3$ and $U_3$ are all order 3 shifts, 
so it is possible to have non-zero relative phases between them. 
The relevant relative phases are $\phi(A_i ,U_3)$, where $i=1,2,3$. 
In the above description of models $S1-4(1,0)$, $F1(1,1)$, $F2(1,1)$ and 
$S1-4$, we have set these phases to be zero. If we let $\phi(A_i ,U_3)$
take non-zero values, {\em i.e.}, $1/3$ or $2/3$, we obtain the same set 
of models. For example, in the construction of the $S1$ model, if we had
taken $\phi(A_3,U_3)\not=0$ instead of $\phi(A_3,U_3)=0$, we would end up 
with the $F1(1,1)$ model. So no new model is generated with this approach.
However, there is another Wilson line, 
\begin{eqnarray}
  A^{\prime}_3=(0,0,0 \vert\vert 0,{\tilde e}^2,{\tilde e}^2)(({1\over 3}
{1\over 3}{1\over 3} {1\over  3}{2\over 3} )^3 \vert 0   )~,\nonumber
\end{eqnarray}
for which new models will emerge. More precisely, if we add this Wilson line 
to the $E1$ and $E2$ models, and set the phase $\phi(A^{\prime}_3,U_3)$ to 
zero, we recover the $S1-4$ models disscussed previously. If we add this 
Wilson line to the $T1(1,1)$ and $T2(1,1)$ models, and set the phase 
$\phi(A^{\prime}_3,U_3)$ to zero, we recover the $F1-4$ models disscussed 
below. If, in the latter case, we take the phase 
$\phi(A^{\prime}_3,U_3)\not=0$,
we recover the models $F1(1,1)$ and $F2(1,1)$. Now, if we add this 
Wilson line to the $E1$ and $E2$ models, and  take the phase 
$\phi(A^{\prime}_3,U_3)\not=0$, we find two new models, which we will 
refer to as $S5(1,1)$ and $S6(1,1)$, respectively. 
Now, let us describe these models.\\
$\bullet$ The $S1(1,1)$ model. Start from the $E1$ model and add the 
$A^{\prime}_3$ Wilson line. Choose the relative phase 
$\phi(A^{\prime}_3,U_3)=1/3$ or $2/3$ (both choices give the same model). 
This model has $SU(2)_1 \otimes SU(6)_3 \otimes U(1)^4$ gauge symmetry. 
The massless spectrum of the $S1(1,1)$ model is given in Table IV. 
Note that the spectrum of this model is very similar to that of the $E1$ model.
In particular, the former can be obtained from the latter via replacing 
$(E_6)_3$ by $SU(6)_3\otimes U(1)$ (the last $U(1)$ normalized to 
$1/\sqrt{6}$ in the first column of Table IV in this paper). Under this 
substitution, ${\bf 78}$ of $(E_6)_3$ is replaced by 
${\bf 35}(0)+{\bf 1}(0)$ of $SU(6)_3 \otimes U(1)$, and ${\bf 27}$ is 
replaced by ${\bf 15}+{\overline {\bf 6}}(+1)+{\overline {\bf 6}}(-1)$. 
(The complex conjugates of these states are substituted similarly.) 
This corresponds to the regular breaking $E_6 \supset SU(6) \otimes U(1)$.\\
$\bullet$ The $S2(1,1)$ model. Start from the $E1$ model and add the 
$A^{\prime}_3$ Wilson line. Choose the relative phase 
$\phi(A^{\prime}_3,U_3)=1/3$ or $2/3$ (both choices give the same model). 
This model has $SU(2)_1 \otimes SU(6)_3 \otimes U(1)^4$ gauge symmetry. 
The massless spectrum of the $S2(1,1)$ model can be obtained from that of 
the $E2$ model by the breaking procedure described for the $S1(1,1)$ model. 
Note that the $S2(1,1)$ model is the same as the $S1(1,1)$ model.\\
Here we note that the $S1(1,1)$ and $S2(1,1)$ models can be obtained from 
the $T1(1,1)$ and $T2(1,1)$ models of Ref\cite{class} by the replacement 
procedure described for the $S2(1,0)$ model. (Similarly, one may construct 
other $SU(6)$ models whose massless spectra would follow from those of the 
corresponding $SO(10)$ models given in Ref\cite{class} by the the 
replacement procedure described for the $S2(1,0)$ model.) \\

{}Next, consider the models $T1(0,0)$ to $T11(0,0)$, and $T1(1,0)$ 
to $T10(1,0)$ of Ref\cite{class}. The following Wilson line can be 
added to all of these models:
\begin{eqnarray}
 A_4=(0,0,0 \vert\vert {\tilde e}^2,0,0)(({1\over 3}{1\over 3}{1\over 3}
{1\over 3}{2\over 3} )^3 \vert 0 )~.\nonumber
\end{eqnarray}
$\bullet$ We will refer to the resulting models as $F1(0,0)$ to $F11(0,0)$
and $F1(1,0)$ to $F10(1,0)$, respectively. These models have $SU(5)_3$ 
grand unified gauge group. 
The massless spectra of these models can be obtained from those of the models 
$T1(0,0)$ to $T11(0,0)$, and $T1(1,0)$ to $T10(1,0)$ by the breaking 
procedure described for the $F1(1,1)$ model. Note that some of these $SU(5)$ 
models are the same just as some of the corresponding $SO(10)$ models are the 
same. For example, $F3(1,0)=F1(1,0)$, which follows from $T3(1,0)=T1(1,0)$. 
Here we note that, in the case of the $T5(1,0)=T6(1,0)$ model, the left-moving 
$SO(8)$ momenta in the $A_4$ Wilson line are given in the $SO(8)\supset SU(3)
\otimes U(1)^2$ basis. In Ref\cite{class} we gave the twists for these 
models with the left-moving $SO(8)$ momenta being written in the 
$SO(8)\supset SU(2)^4$ basis. It is not difficult to translate between 
these two bases. 
In particular, the $A_4$ Wilson line in the latter basis reads:
\begin{eqnarray}
 A_4=(0,0,0 \vert\vert {\tilde e}^2\vert 0,0,0,0)
 (({1\over 3}{1\over 3}{1\over 3}{1\over 3}{2\over 3} )^3
 \vert 0 )~.\nonumber
\end{eqnarray}
In Ref\cite{class} it was pointed out that the models $T1(1,0)$, $T2(1,0)$,
$T3(1,0)$ and $T4(1,0)$ can be obtained via assymmetric orbifolds. 
Here we will give these asymmetric orbifolds once again as they will be 
important for constructing new $SU(5)$ models:\\
$\bullet$ The $T1(1,0)$ model. Start from the $N1(1,0)$ model of 
Ref \cite{class} and perform the following twists:
\begin{eqnarray}
 &&T_3 =(\theta,\theta,\theta\vert\vert 0, e_1/3,  \theta)
 ({\cal P} \vert  2/3)~,\nonumber\\
 &&T_2=(0, \sigma,\sigma\vert\vert  e_1/2, e_1/2, 0)
 (0^{15} \vert 0)~.\nonumber
\end{eqnarray}  
This model has $SU(2)_1 \otimes SU(2)_3 \otimes SO(10)_3 \otimes U(1)^3$ 
gauge symmetry. The massless spectrum of the $T1(1,0)$ model is given in 
Table VI of Ref\cite{class}. Note that this construction is different 
from the one presented in Sec. III. The equivalence of these two 
constructions are thoroughly explained in Ref\cite{kt}. \\
$\bullet$ The $T2(1,0)$ model. Start from the $N1(1,0)$ model of 
Ref\cite{class} and perform the following twists:
\begin{eqnarray}
 &&T_3 =(\theta,\theta,\theta\vert\vert e_1/3, 0, \theta)
 ({\cal P} \vert  2/3)~,\nonumber\\
 &&T_2=(0, \sigma,\sigma\vert\vert  e_1/2, e_1/2,0)
 (0^{15} \vert 0)~.\nonumber
\end{eqnarray}  
This model has $SU(2)_1 \otimes SU(2)_3 \otimes SO(10)_3 \otimes U(1)^3$ 
gauge symmetry. The massless spectrum of the $T2(1,0)$ model is given in 
Table VI of Ref \cite{class}.\\
$\bullet$ The $T3(1,0)$ model. Start from the $N1(1,0)$ model of 
Ref \cite{class} and perform the following twists:
\begin{eqnarray}
 &&T_3 =(\theta,0,\theta\vert\vert e_1/3, 0, \theta)
 ({\cal P} \vert  2/3)~,\nonumber\\
 &&T_2=(0, \sigma,\sigma\vert\vert  e_1/2, e_1/2, 0)
 (0^{15} \vert 0)~.\nonumber
\end{eqnarray}  
This model has $SU(2)_1 \otimes SU(2)_3 \otimes SO(10)_3 \otimes U(1)^3$ 
gauge symmetry. The massless spectrum of the $T3(1,0)$ model is given in 
Table VII of Ref\cite{class}.\\
$\bullet$ The $T4(1,0)$ model. Start from the $N1(1,0)$ model of 
Ref \cite{class} and perform the following twists:
\begin{eqnarray}
 &&T_3 =(\theta,0,\theta\vert\vert  0, e_1/3, \theta)
 ({\cal P} \vert  2/3)~,\nonumber\\
 &&T_2=(0, \sigma,\sigma\vert\vert  e_1/2, e_1/2, 0)
 (0^{15} \vert 0)~.\nonumber
\end{eqnarray}  
This model has $SU(2)_1 \otimes SU(2)_3 \otimes SO(10)_3 \otimes U(1)^3$ 
gauge symmetry. The massless spectrum of the $T4(1,0)$ model is given in 
Table VII of Ref \cite{class}.\\
$\bullet$ The $T7-10(1,0)$ models. If, instead of the $N1(1,0)$ model, 
we start from the $N2(1,0)$ model and perform the 
above twists for the $T1-4(1,0)$ models, we obtain the $T7-10(1,0)$ 
models of Ref\cite{class}.
The models $T1-4(1,0)$ and $T7-10(1,0)$ now admit two more Wilson lines:
\begin{eqnarray}
 &&A_5=(0,0,0 \vert\vert 0, {\tilde e}^2,0)(({1\over 3}{1\over 3}{1\over 3}
{1\over 3}{2\over 3} )^3 \vert 0 )~,\nonumber\\
 &&A_6=(0,0,0 \vert\vert {\tilde e}^2, -{\tilde e}^2,0)(({1\over 3}{1\over 3}
{1\over 3} {1\over 3}{2\over 3} )^3 \vert 0 )~.\nonumber
\end{eqnarray}
{}Adding these Wilson lines, we obtain the following $SU(5)$ models:\\
$\bullet$ The $F11(1,0)$ model. Start from either the $T1(1,0)$ or $T2(1,0)$ 
model (both choices give the same model) and add the $A_5$ Wilson line. 
This model has $SU(2)_1 \otimes SU(2)_3 \otimes SU(5)_3 \otimes U(1)^4$ 
gauge symmetry. 
The massless spectrum of the $F11(1,0)$ model is given in Table IV.  
Note that the spectrum of this model is very similar to that of the $E1$ model.
In particular, the former can be obtained from the latter via replacing 
$(E_6)_3$ by $SU(5)_3 \otimes SU(2)_3 \otimes U(1)$ (the last $U(1)$ 
normalized to $1/\sqrt{6}$ in the second column of Table IV in this paper). 
Under this substitution, ${\bf 78}$ of  $(E_6)_3$ is replaced by 
$({\bf 24},{\bf 1})(0)+({\bf  1},{\bf 3})(0)+({\bf 1},{\bf 1})(0)$ of 
$SU(5)_3 \otimes SU(2)_3 \otimes U(1)$, and ${\bf 27}$ is replaced by 
$({\bf 10},{\bf 1})(-2)+({\bf  5},{\bf 1})(+4)+
({\overline {\bf 5}},{\bf 2})(+1)+({\bf  1},{\bf 2})(-5)$. 
(The complex conjugates of these states are substituted similarly.) 
This corresponds to the regular breaking $E_6 \supset SU(5) \otimes SU(2) 
\otimes U(1)$. \\
$\bullet$ The $F12(1,0)$ model. Start from either the $T3(1,0)$ or $T4(1,0)$ 
model (both choices give the same model) and add the $A_5$ Wilson line. 
This model has $SU(2)_1 \otimes SU(2)_3 \otimes SU(5)_3 \otimes U(1)^4$ 
gauge symmetry. 
The massless spectrum of the $F12(1,0)$ model can be obtained from that of 
the $E2$ model by the breaking procedure described for the $F11(1,0)$ model. 
Note that $F12(1,0) = F11(1,0)$.\\
$\bullet$ The $F13(1,0)$ model. Start from either the $T7(1,0)$ or $T8(1,0)$ 
model (both choices give the same model) and add the $A_5$ Wilson line. 
This model has $SU(2)_1 \otimes SU(2)_3 \otimes SU(5)_3 \otimes U(1)^4$ 
gauge symmetry. The massless spectrum of the $F13(1,0)$ model is the same 
as that of the $F11(1,0)$ model with additional states coming from the $T_2$ 
sector. These states read: $4({\bf 2},{\bf 1},{\bf 1})(0,\pm 3,0,0)_L$.\\ 
$\bullet$ The $F14(1,0)$ model. Start from either the $T9(1,0)$ or $T10(1,0)$ 
model (both choices give the same model) and add the $A_5$ Wilson line. 
This model has $SU(2)_1 \otimes SU(2)_3 \otimes SU(5)_3 \otimes U(1)^4$ 
gauge symmetry. The massless spectrum of the $F14(1,0)$ model is the same 
as that of the $F12(1,0)$ model with additional states coming from the 
$T_2$ sector. These states read: 
$4({\bf 2},{\bf 1},{\bf 1})(0,\pm 3,0,0)_L$. 
Note that $F14(1,0) = F13(1,0)$. \\ 
$\bullet$ The $F1$ model. Start from either the $T1(1,0)$ or $T2(1,0)$ model 
(both choices give the same model) and add the $A_6$ Wilson line. 
Choose the relative phase between the $T_3$ and $A_6$  twisted sectors to 
be $\phi(T_3, A_6)=0$ or $1/3$ (both choices give the same model). 
This model has $SU(3)_1 \otimes SU(5)_3 \otimes U(1)^4$ gauge symmetry. 
The massless spectrum of the $F1$ model is given in Table V.\\
$\bullet$ The $F2$ model. Start from either the $T1(1,0)$ or $T2(1,0)$ model 
(both choices give the same model) and add  the $A_6$ Wilson line. 
Choose the relative phase between the $T_3$ and $A_6$  twisted sectors to 
be $\phi(T_3, A_6)=2/3$. This model has 
$SU(2)_1 \otimes SU(2)_1 \otimes SU(5)_3 \otimes U(1)^4$ gauge symmetry. 
The massless spectrum of the $F2$ model is given in Table V.\\
$\bullet$ The $F3$ model. Start from either the $T3(1,0)$ or $T4(1,0)$ model 
(both choices give the same model) and add  the $A_6$ Wilson line. 
Choose the relative phase between the $T_3$ and $A_6$  twisted sectors to 
be $\phi(T_3, A_6)=0$ or $1/3$ (both choices give the same model). 
This model has $SU(3)_1 \otimes SU(5)_3 \otimes U(1)^4$ gauge symmetry. 
The massless spectrum of the $F3$ model is given in Table VI. 
Note that $F3 = F1(1,0)$. \\
$\bullet$ The $F4$ model. Start from either the $T3(1,0)$ or $T4(1,0)$ 
model (both choices give the same model) and add  the $A_6$ Wilson line. 
Choose the relative phase between the $T_3$ and $A_6$ twisted sectors to 
be $\phi(T_3, A_6)=2/3$. This model has 
$SU(2)_1 \otimes SU(2)_1 \otimes SU(5)_3 \otimes U(1)^4$ gauge symmetry. 
The massless spectrum of the $F4$ model is given in Table VI. 
Note that $F4 = F2(1,0)$. \\
$\bullet$ The $F5$ model. Start from either the $T7(1,0)$ or $T8(1,0)$ model 
(both choices give the same model) and add  the $A_6$ Wilson line. 
Choose the relative phase between the $T_3$ and $A_6$ twisted sectors to be 
$\phi(T_3, A_6)=0$ or $1/3$ (both choices give the same model). 
This model has $SU(3)_1 \otimes SU(5)_3 \otimes U(1)^4$ gauge symmetry. 
The massless spectrum of the $F5$ model is the same as that of the $F1$ model 
with additional states coming from the $T_2$ sector. 
These additional states are given in Table VII.\\
$\bullet$ The $F6$ model. Start from either the $T7(1,0)$ or $T8(1,0)$ model 
(both choices give the same model) and add  the $A_6$ Wilson line. 
Choose the relative phase between the $T_3$ and $A_6$  twisted sectors to 
be $\phi(T_3, A_6)=2/3$. This model has 
$SU(2)_1 \otimes SU(2)_1 \otimes SU(5)_3 \otimes U(1)^4$ gauge symmetry. 
The massless spectrum of the $F6$ model is the same as that of the $F2$ model 
with additional states coming from the $T_2$ sector. 
These additional states are given in Table VII.\\
$\bullet$ The $F7$ model. Start from either the $T9(1,0)$ or $T10(1,0)$ model 
(both choices give the same model) and add  the $A_6$ Wilson line. 
Choose the relative phase between the $T_3$ and $A_6$ twisted sectors to be 
$\phi(T_3, A_6)=0$ or $1/3$ (both choices give the same model). 
This model has $SU(3)_1 \otimes SU(5)_3 \otimes U(1)^4$ gauge symmetry. 
The massless spectrum of the $F7$ model is the same as that of the $F3$ model 
with additional states coming from the $T_2$ sector. These additional states 
are given in Table VII. 
Note that $F7 = F5$.\\
$\bullet$ The $F8$ model. Start from either the $T9(1,0)$ or $T10(1,0)$ model 
(both choices give the same model) and add  the $A_6$ Wilson line. 
Choose the relative phase between the $T_3$ and $A_6$  twisted sectors to be 
$\phi(T_3, A_6)=2/3$. This model has 
$SU(2)_1 \otimes SU(2)_1 \otimes SU(5)_3 \otimes U(1)^4$ gauge symmetry. 
The massless spectrum of the $F8$ model is the same as that of the $F4$ 
model with additional states coming from the $T_2$ sector. 
These additional states are given in Table VII.
Note that $F8 = F6$. \\

{}Next, consider the $T5(1,0)$ and $T6(1,0)$ models of Ref\cite{class}. 
These models admit the $A_4$ Wilson line. 
The resulting models are $F5(1,0)$ and $F6(1,0)$, as we disscussed previously.
There are, however, additional Wilson lines that can be added to these 
two models. The $T6(1,0)$ model admits the following four Wilson lines 
(here the left-moving $SO(8)$ momenta are given in 
the $SO(8)\supset SU(2)^4$ basis):
\begin{eqnarray}
 &&B_1=(0,0,0 \vert\vert {\tilde e}^2 \vert, 0,\sqrt{2}/3,\sqrt{2}/3, 
-\sqrt{2}/3)               (({1\over 3}{1\over 3}{1\over 3}{1\over 3}{2\over 3} )^3 \vert 0 )~,\nonumber\\
 &&B_2=(0,0,0 \vert\vert 0 \vert, 0,\sqrt{2}/3,\sqrt{2}/3, -\sqrt{2}/3)
                 (({1\over 3}{1\over 3}{1\over 3}{1\over 3}{2\over 3} )^3 \vert 0 )~,\nonumber\\
 &&B_3=(0,0,0 \vert\vert e_1/3 \vert -\sqrt{2}/3,\sqrt{2}/3,0,0)(({1\over 3}{1\over 3}{1\over 3}
                 {1\over 3}{2\over 3} )^3 \vert 0 )~,\nonumber\\
 &&B_4=(0,0,0 \vert\vert e_1/3 \vert 0,0,\sqrt{2}/3,-\sqrt{2}/3)(({1\over 3}{1\over 3}{1\over 3}
                 {1\over 3}{2\over 3} )^3 \vert 0 )~.\nonumber
\end{eqnarray}
The $T5(1,0)$ model admits the Wilson lines $B_1$, $B_2$ and $B_3$, but 
not $B_4$. {\em Apriori}, there are three cases to consider for each 
Wilson line, namely, there are three choices of the relative phases 
$\phi(T_3,B_i)$. For the sake of brevity, we will not list all the possible 
models. However, let us consider one of these models for illustration. \\
$\bullet$ The $F9$ model. Start from the $T6(1,0)$ model and add the 
$B_1$ Wilson line. Choose the relative phase between the $T_3$ and $B_1$ 
sectors to be $\phi(T_3,B_1)=2/3$. 
This model has $SU(4)_1 \otimes SU(2)_1 \otimes SU(5)_1 \otimes U(1)^4$ gauge
symmetry. The massless spectrum of this model is given in Table VIII. 
Note that $SU(4)_1$ is the hidden sector gauge group in this model, 
whereas $SU(2)_1$ is the horizontal symmetry gauge group. 
Also note that this model is completely anomaly-free.

{}Here we note that some of the models obtained by adding the above 
$B_i$ Wilson lines to the $T5(1,0)$ and $T6(1,0)$ models do not have 
any hidden sector, although some of them do possess non-abelian horizontal 
gauge symmetries (these are products of $SU(2)_1$). 
Examples of such models are the following. 
Start from the $T5(1,0)$ model and add either the $B_1$ or $B_2$ Wilson line. 
Then, regardless of the choice of $\phi(T_3,B_i)$, the resulting model does 
not have a hidden sector. Another point that we should stress is that some 
of the models have anomalous $U(1)$s in their spectra. 
An example of this is the following. Start from the $T6(1,0)$ model and 
add the $B_1$ Wilson line. Choose the relative phase between the $T_3$ and 
$B_1$ sectors to be $\phi(T_3,B_1)=0$ or $2/3$ 
(both choices give the same model). 
This model has $SU(2)_1 \otimes SU(2)_1 \otimes SU(5)_1 \otimes U(1)^6$ gauge 
symmetry. Here $SU(2)_1\otimes SU(2)_1$ is the hidden sector gauge group. 
This model has an anomalous $U(1)$.
The generic feature of all of these models is that they are $SU(5)_3$ grand 
unified string models with three families of quarks and leptons, and they 
have only one adjoint (and no other higher dimensional representation) Higgs 
field, and the adjoint Higgs field does not carry any gauge quantum numbers 
other than those of the grand unified gauge group. One adjoint Higgs field 
neutral under other gauge quantum numbers is a generic feature for all 
three-family models constructed in this paper and in Ref\cite{class}.

{}Finally, let us summarize the new versus old results presented in this 
paper. The models $F1(1,1)$ and $S2(1,0)$ were first presented in 
Ref\cite{kt}. There the possibility of having an enhanced hidden sector 
gauge group was also pointed out. The first examples of models with 
enhanced hidden sectors were constructed in Ref\cite{five}. 
In particular, Ref\cite{five} presented the models $F1$, $F2$, and 
briefly discussed the $S1$ and $S2$ models. The rest of the models 
classified in this paper are new. This concludes the construction of 
three-family $SU(5)$ and $SU(6)$ string models. Together with 
Ref\cite{class}, the present paper completes the construction of 
three-family grand unification in string theory.  

\section{Moduli Space of $SU(5)$ and $SU(6)$ Models}
\bigskip

{}In Ref\cite{class}, we pointed out that all but one of the 
$SO(10)$ models constructed there are connected by flat moduli. 
Such connections also exist for most of the $SU(5)$ and $SU(6)$ models 
constructed in this paper as well. Let us collect them into three sets: 

({\em i}) Starting from Fig.1 of Ref\cite{class}, the same picture will 
hold if we replace all the $SO(10)$ models there by the corresponding 
$SU(5)$ models, and the 
$E1 = E2$ model by the $T1(1,1) = T2(1,1)$ model, respectively, 

({\em ii}) Starting from Fig.1 of Ref\cite{class}, the same picture will
hold if we replace all the $SO(10)$ models there by the corresponding
$SU(6)$ models, while keeping the $E1 = E2$ model untouched.
Since the $E1 = E2$ model is connected to the $T1(1,1) = T2(1,1)$ model,
these two sets are connected as well.

({\em iii}) The rest of the connections are summarized in Fig.1 in this 
paper. There, the symbols $a$, $b$, $c$, $d$, $e$, $f$, $g$ stand for 
the fields whose vevs are the corresponding flat directions.
Thus, $a$ corresponds to the adjoint breaking of $SO(10)$ down to 
$SU(5)\otimes U(1)$; $b$ and $g$ stand for vevs of singlets, {\em i.e.},
 neutral under all the gauge symmetries; 
$c$ corresponds to the adjoint breaking of $E_6$ down to 
$SU(5)\otimes SU(2) \otimes U(1)$; $d$ stands for the triplet Higgs 
({\em i.e.}, $({\bf 1},{\bf 3},{\bf 1})(0,0,0)_L$) of the $SU(2)_3$ subgroup 
in models $S1(1,0)=S3(1,0)$ and $S2(1,0)=S4(1,0)$; 
$e$ corresponds to the adjoint breaking of $SU(6)$ down to 
$SU(5)\otimes U(1)$; finally, $f$ stands for the adjoint breaking of 
$E_6$ down to $SU(6)\otimes U(1)$.

This completes our discussion of the moduli space of $3$-family
$SU(5)$ and $SU(6)$ string models constructed in this paper,
and their connections to the $SO(10)$ and the $E_6$ models.

\section{Remarks}
\bigskip

Together, Ref \cite{class} and this paper give the classification of all
${\bf Z}_6$ asymmetric orbifolds that can yield $3$-family supersymmetric 
grand unified string models that have a non-abelian hidden sector.
The massless spectra of many of these models
are presented in the tables in these two papers.
Note that the number of models increases as the rank of the GUT gauge 
symmetry decreases: there is a unique $E_6$ model, but there are dozens
of $SU(5)$ models. This is expected, since a smaller GUT symmetry 
permits more room for the hidden and the intermediate sectors, so the 
number of choices for the latter increases. 

It remains an open question if this is a complete classification of all
$3$-family GUT models within the framework of conformal field theory and 
asymmetric orbifolds. One possibility that is outside of our framework is the
the non-diagonal embedding of the $SU(5)_3$ current algebra. Although we 
have argued that it is unlikely that other higher-level realizations
of the current algebras can yield interesting $3$-family GUT models, it 
will be important to explore them as well. Presumably, $3$-family GUT 
string models can also be obtained in non-perturbative string theory.
This direction requires a better understanding of string theory.

Phenomenologically, are any of the $3$-family GUT string models constructed 
so far viable? 
There are many issues that one has to address before one can come to
some sort of an answer to this question. The first step is clearly to
determine the couplings of any specific model. This is certainly within 
the reach of our present understanding. Knowing the superpotential, or at 
least some of its leading terms, will definitely help. This knowledge may 
also shed light on how these GUT models differ from the more conventional 
standard model strings.

\acknowledgments

\bigskip

{}We would like to thank Michael Bershadsky, Pran Nath, 
Gary Shiu, Tom Taylor and Yan Vtorov-Karevsky for discussions.
The research of S.-H.H.T. was partially supported by National Science 
Foundation.
The work of Z.K. was supported in part by the grant NSF PHY-96-02074, and 
the DOE 1994 OJI award. Z.K. would also like to thank Mr. Albert Yu 
and Mrs. Ribena Yu for financial support.

%%%%%%%%%%%%%%%%%%%%%%%%%%%%%%%%%%%%%%%%%%%%%%%%
\newpage
\begin{figure}[t]
%\hspace*{}
%\vspace*{}
\epsfxsize=16 cm
\epsfbox{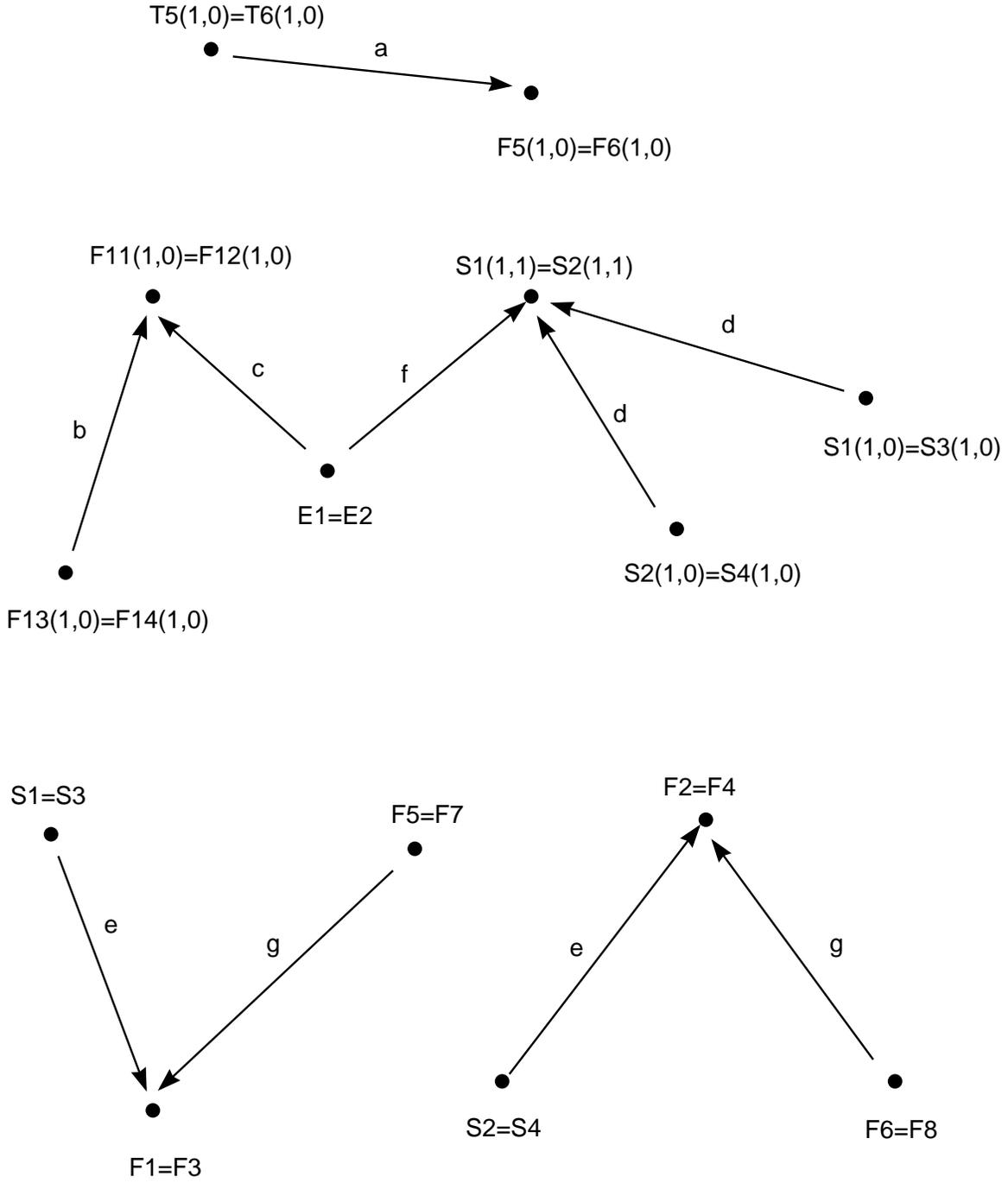}
\caption{The arrows correspond to flat directions. 
Here $a$, $b$, $c$, $d$, $e$, $f$ and $g$ stand for the fields whose vevs are the 
corresponding flat directions. See section V for details.}
\end{figure}
%%%%%%%%%%%%%%%%%%%%%%%%%%%%%%%%%%%%%%%%%%%%%%%%

%%%%%%%%%%%%%%%%%% TABLE I %%%%%%%%%%%%%%%%%%%%%%
%%%%%%%%%%%%%%%%%%%%%%%%%%%%%%%%%%%%%%%%%%%%%%%%%%%%%%%%%%%%%%%%%%%%%%%%%%%%%%%
\begin{table}[t]
\begin{tabular}{|c|l|l|}
%%%%%%%%%%%%%%%%%%%%%%%%%%%%%%%%%%%%%%%%%%%%%%%%%%%%%%%%%%%%%%%%%%%%%%%%%%%%%%%
  & $S2(1,0)$ & $F1(1,1)$ \\
 M & $SU(2)^2 \otimes SU(6) \otimes U(1)^3$ 
      &  $SU(2) \otimes SU(5) \otimes U(1)^5$ \\ \hline
%%%%%%%%%%%%%%%%%%%%%%%%%%%%%%%%%%%%%%%%%%%%%%%%%%%%%%%%%%%%%%%%%%%%%%%%%%%%%%%
  & & \\
       & $ ({\bf 1},{\bf 1},{\bf 35})(0,0,0)_L$
    & $ ({\bf 1},{\bf 24})(0,0,0,0,0)_L$ \\
  $U$ 
    & $2 ({\bf 1},{\bf 1},{\bf 1})(-{3},\pm 3,0)_L$
    & $2 ({\bf 1},{\bf 1})(0,-{3},\pm 3,0,0)_L$ \\
    &  $ ({\bf 1},{\bf 1},{\bf 1})(+6,0,0)_L$
    & $ ({\bf 1},{\bf 1})(0,+6,0,0,0)_L$ \\
    &  $ 2({\bf 2},{\bf 2},{\bf 1})(0,0,0)_L$ 
   & $2 ({\bf 1},{\bf 1})(0,0,0,0,0)_L$ \\
   & $({\bf 1},{\bf 3},{\bf 1})(0,0,0)_L$ & \\
\hline
%%%%%%%%%%%%%%%%%%%%%%%%%%%%%%%%%%%%%%%%%%%%%%%%%%%%%%%%%%%%%%%%%%%%%%%%%%%%%%%
%%%%%%%%%%%%%%%%%%%%%%%%%%%%%%%%%%%%%%%%%%%%%%%%%%%%%%%%%%%%%%%%%%%%%%%%%%%%%%%
 & & \\
    & $ ({\bf 1},{\bf 1},{\overline {\bf 6}})(-{2},0,\pm {1})_L$
    & $ ({\bf 1},{\overline {\bf 5}})(0,-{2},0, -{1},+3)_L$  \\
    & $ ({\bf 1},{\bf 1},{\bf 15})(-2,0,0)_L$
    & $ ({\bf 1},{\bf 10})(0,-2,0,-1,-1)_L$ \\
    & $2 ({\bf 1},{\bf 1},{\overline {\bf 6}})(+1,\pm 1,\pm 1)_L$
    & $ ({\bf 1},{\bf 1})(0,-2,0,-1,-5)_L$ \\
    & $2 ({\bf 1},{\bf 1},{\bf 15})(+1,\pm 1,0)_L$ 
    &  $ ({\bf 1},{\bf 5})(0,-2,0,+2,+2)_L$ \\
 $T3$ &  & $  ({\bf 1},{\overline {\bf 5}})(0,-2,0,+{2},-2)_L$ \\
   &  &  $ ({\bf 1},{\bf 1})(0,-2,0,-4,0)_L$ \\
   & & $2 ({\bf 1},{\overline {\bf 5}})(0,+1,\pm 1,-1,+3)_L$  \\
   & & $2 ({\bf 1},{\bf 10})(0,+1,\pm 1,-1,-1)_L$ \\
   & & $2 ({\bf 1},{\bf 1})(0,+1,\pm 1,-1,-5)_L$ \\
   & & $2 ({\bf 1},{\bf 5})(0,+1, \pm 1,+2,+2)_L$ \\
   & & $2 ({\bf 1},{\overline {\bf 5}})(0,+1,\pm 1,+{2},-2)_L$ \\
   & & $2 ({\bf 1},{\bf 1})(0,+1, \pm 1,-4,0)_L$ \\
   & & \\
 \hline
%%%%%%%%%%%%%%%%%%%%%%%%%
  & &\\
       & $ ({\bf 1}, {\bf 2},{\bf 6}) (-{1},0, \pm 1)_L$
    & $ ({\bf 1},{\bf 5}) (\pm 1,-{1},0,+1,-3)_L$ \\
   &  $ ({\bf 1},{\bf 2},{\overline {\bf 15}})(-{1},0,0)_L$
      &  $ ({\bf 1},{\overline {\bf 10}})(\pm 1,-1,0,+1,+1)_L$ \\
    $T6$ & & $({\bf 1},{\bf 1})(\pm 1,-1,0,+1,+5)_L$ \\
   & & $({\bf 1},{\bf 5})(\pm 1,-1,0,-2,+2)_L$ \\
   & & $ ({\bf 1},{\overline {\bf 5}})(\pm 1,-1,0,-{2},-2)_L$ \\
   & & $ ({\bf 1},{\bf 1})(\pm 1,-1,0,+4,0)_L$ \\
 & & \\
 \hline
%%%%%%%%%%%%%%%%%%%%%%%%%%%%%%%%%%%%%%%%%%%%%%%%%%%%%%%%%%%%%%%%%%%%%%%%%%%%%%%
 & & \\
   $T2$     & $({\bf 2},{\bf 1},{\bf 1})(0,{\pm 3},0)_L$
    & $({\bf 2},{\bf 1})(0,0,{\pm 3},0,0)_L$ \\
    & $({\bf 1},{\bf 4},{\bf 1})(+{3},0,0)_L$
    & $({\bf 1},{\bf 1})(\pm {3},+{3},0,0,0)_L$ \\
  & $2({\bf 1},{\bf 2},{\bf 1})(-3,0,0)_L$ & \\
  & $({\bf 1},{\bf 2},{\bf 1})(+3,0,0)_L$ & \\
\hline
%%%%%%%%%%%%%%%%%%%%%%%%%%%%%%%%%%%%%%%%%%%%%%%%%%%%%%%%%%%%%%%%%%%%%%%%%%%%%%%
 & & \\
 $U(1)$    & $(1/{3\sqrt{2}}, 1/\sqrt{6}, 1/\sqrt{6})$
   & $(1/\sqrt{6}, 1/{3\sqrt{2}},1/\sqrt{6},1/{6},1/{2\sqrt{15}})$ \\
%%%%%%%%%%%%%%%%%%%%%%%%%%%%%%%%%%%%%%%%%%%%%%%%%%%%%%%%%%%%%%%%%%%%%%%%%%%%%%%
\end{tabular}
%%%%%%%%%%%%%%%%%%%%%%%%%%%%%%%%%%%%%%%%%%%%%%%%%%%%%%%%%%%%%%%%%%%%%%%%%%%%%%%
\caption{The massless spectra of the $S2(1,0)$ and $F1(1,1)$ models
with gauge groups
$SU(2)_1 \otimes SU(2)_3 \otimes SU(6)_3 \otimes U(1)^3$
and (3) $SU(2)_1 \otimes SU(5)_3 \otimes U(1)^5$, respectively.
The $U(1)$ normalization
radii are given at the bottom of the Table.
The gravity, dilaton and gauge supermultiplets are not shown.}
\end{table}
%%%%%%%%%%%%%%% END OF TABLE I %%%%%%%%%%%%%%%%%%%%%%%%
%%%%%%%%%%%%%%%%%%%%%%%%%%%%%%%%%%%%%%%%

%%%%%%%%%%%%%%%%TABLE II %%%%%%%%%%%%%%%%%%%%%%%%%%%
%%%%%%%%%%%%%%%%%%%%%%%%%%%%%%%%%%%%%%%%%%%%%%%%%%%%%%%%%%%%%%%
\begin{table}[t]
\begin{tabular}{|c|l|l|} 
%%%%%%%%%%%%%%%%%%%%%%%%%%%%%%%%%%%%%%%%%%%%%%%%%%%%%%%%%%%%%%%%%%%%%%%%%%%%%%%
M    &  $S1$  & $S2$ \\
       & $SU(3)   \otimes SU(6) \otimes U(1)^3$ &
    $SU(2)^2 \otimes SU(6) \otimes U(1)^3$ \\
   \hline
%%%%%%%%%%%%%%%%%%%%%%%%%%%%%%%%%%%%%%%%%%%%%%%%%%%%%%%%%%%%%%%%%%%%%%%%%%%%%%%
 & & \\  
 & $({\bf 1},{\bf 35})(0,0,0)_L$
   & $ ({\bf 1},{\bf 1},{\bf 35})(0,0,0)_L$ \\
 &  $({\bf 1},{\bf 1})(0,0,0)_L$
   & $ ({\bf 1},{\bf 1},{\bf 1})(0,0,0)_L$ \\
 $U$    &  $ ({\bf 1},{\bf 1}) (+6,0,0)_L$ 
   & $ ({\bf 1},{\bf 1},{\bf 1})(0,0,-6)_L$ \\
    & $2 ({\bf 1},{\bf 1})(-3, {\pm 3},{\pm 3})_L$ &
     $2 ({\bf 1},{\bf 2},{\bf 1})({\pm 1},{\mp 3},+3)_L$ \\
    & $2 ({\overline {\bf 3}},{\bf 1})(+3,-3,+1)_L$ & 
     $  ({\bf 2},{\bf 1}, {\bf 1})({\pm 2},0,+3)_L$\\
   & $({\bf 3},{\bf 1})(0,0,-4)_L$ & \\
  & & \\
 \hline
%%%%%%%%%%%%%%%%%%%%%%%%%%%%%%%%%%%%%%%%%%%%%%%%%%%%%%%%%%%%%%%%%%%%%%%%%%%%%%%
%%%%%%%%%%%%%%%%%%%%%%%%%%%%%%%%%%%%%%%%%%%%%%%%%%%%%%%%%%%%%%%%%%%%%%%%%%%%%%%
   & & \\
    & $6 ({\bf 1},{\bf 15})(+1,-1,-{1})_L$ &
     $3 ({\bf 1},{\bf 1},{\bf 15})(0,0,+2)_L$ \\
  $T3$ &       $6 ({\bf 1},{\overline {\bf 6}})(-2,+1,-{1})_L$ & 
     $6 ({\bf 1},{\bf 1},{\overline {\bf 6}})(\pm 1,\mp {1},-1)_L$ \\
    &  $3 ({\bf 1},{\overline {\bf 6}})(+1,0,+2)_L$ & \\ 
  & & \\
   \hline
%%%%%%%%%%%%%%%%%%%%%%%%%%%%%%%%%%%%%%%%%%%%%%%%%%%%%%%%%%%%%%%%%%%%%%%%%%%%%%%
 & & \\ 
 $T6$ &       $3({\bf 1},{\bf 6})(+2,+1,+1)_L$ &
      $ 3({\bf 1},{\bf 1}, {\bf 6})(\pm 1,\pm {1},+1)_L$ \\
  & $ 3({\bf 1},{\overline {\bf 15}}) ( -1,-{1},+{1})_L$ & \\
  & & \\
   \hline 
%%%%%%%%%%%%%%%%%%%%%%%%%%%%%%%%%%%%%%%%%%%%%%%%%%%%%%%%%%%%%%%%%%%%%%%%%%%%%%%
 & & \\   
    &  $({\bf 3},{\bf 1})(\pm 3,-3,-1)_L$ &
  $({\bf 2},{\bf 2},{\bf 1})(\pm 1,\mp 3,0)_L$ \\
 $T2$   &  $({\overline {\bf 3}},{\bf 1})(-3,+3,+1)_L$ &
 $({\bf 1},{\bf 2},{\bf 1})(\pm 1,\pm 3, -3)_L$ \\
 & $ ({\bf 1},{\bf 1})(+3,\pm 3,\mp 3)_L$ & \\
 & & \\
 \hline
%%%%%%%%%%%%%%%%%%%%%%%%%%%%%%%%%%%%%%%%%%%%%%%%%%%%%%%%%%%%%%%%%%%%%%%%%%%%%%%
 $U(1)$  &
    $({1\over{3\sqrt{2}}},~{1\over{2\sqrt{3}}},~{1\over 
    {2\sqrt{3}}})$ &
  $({1\over 2},~{1\over {2\sqrt{3}}},~{1\over {3\sqrt{2}}})$~ \\
%%%%%%%%%%%%%%%%%%%%%%%%%%%%%%%%%%%%%%%%%%%%%%%%%%%%%%%%%%%%%%%%%%%%%%%%%%%%%%%
\end{tabular}
%%%%%%%%%%%%%%%%%%%%%%%%%%%%%%%%%%%%%%%%%%%%%%%%%%%%%%%%%%%%%%%%%%%%%%%%%%%%%%%
\caption{The massless spectra of the two $SU(6)$ models $S1$ and 
$S2$ with gauge symmetries 
$SU(3)_1 \otimes SU(6)_3 \otimes U(1)^3$ and 
$SU(2)_1 \otimes SU(2)_1 \otimes SU(6)_3 \otimes U(1)^3$, respectively. 
Note that double signs (as in $({\bf 1},{\bf 2},{\bf 1})(\pm 1,\pm 3, -3)_L$)
are correlated. The $U(1)$ normalization 
radii are given at the bottom of the table.
The graviton, dilaton and gauge supermultiplets are not shown.}

\end{table}
%%%%%%%%%%%%%%%%%END OF TABLE II %%%%%%%%%%%%%%%%%%%
%%%%%%%%%%%%%%%%%%%%%%%%%%%%%%%%%%%%%%%%%%%%%%%%%%%%%%%%%%%%%%

%%%%%%%%%%%%%%%%TABLE III%%%%%%%%%%%%%%%%%%%%%%%%%%%
%%%%%%%%%%%%%%%%%%%%%%%%%%%%%%%%%%%%%%%%%%%%%%%%%%%%%%%%%%%%%%%
\begin{table}[t]
\begin{tabular}{|c|l|l|} 
%%%%%%%%%%%%%%%%%%%%%%%%%%%%%%%%%%%%%%%%%%%%%%%%%%%%%%%%%%%%%%%%%%%%%%%%%%%%%%%
M    &  $S3$  & $S4$ \\
       & $SU(3)   \otimes SU(6) \otimes U(1)^3$ &
    $SU(2)^2 \otimes SU(6) \otimes U(1)^3$ \\
   \hline
%%%%%%%%%%%%%%%%%%%%%%%%%%%%%%%%%%%%%%%%%%%%%%%%%%%%%%%%%%%%%%%%%%%%%%%%%%%%%%%
 & & \\  
 & $({\bf 1},{\bf 35})(0,0,0)_L$
   & $ ({\bf 1},{\bf 1},{\bf 35})(0,0,0)_L$ \\
 &  $({\bf 1},{\bf 1})(0,0,0)_L$
   & $ ({\bf 1},{\bf 1},{\bf 1})(0,0,0)_L$ \\
 $U$    &  $ ({\bf 1},{\bf 1}) (+6,0,0)_L$ 
   & $ ({\bf 1},{\bf 1},{\bf 1})(0,0,-6)_L$ \\
  &  $({\overline {\bf 3}},{\bf 1})(-3,+3,+1)_L$ &
 $({\bf 1},{\bf 2},{\bf 1})(\pm 1,\pm 3, -3)_L$ \\
    &     $ ({\bf 1},{\bf 1})(+3,\pm 3,\mp 3)_L$                   & 
     $  ({\bf 2},{\bf 1}, {\bf 1})({\pm 2},0,+3)_L$\\
  &  $({\bf 3},{\bf 1})(\pm 3,-3,-1)_L$ &
  $({\bf 2},{\bf 2},{\bf 1})(\pm 1,\mp 3,0)_L$ \\
   & $({\bf 3},{\bf 1})(0,0,-4)_L$ & \\
  & & \\
 \hline
%%%%%%%%%%%%%%%%%%%%%%%%%%%%%%%%%%%%%%%%%%%%%%%%%%%%%%%%%%%%%%%%%%%%%%%%%%%%%%%
%%%%%%%%%%%%%%%%%%%%%%%%%%%%%%%%%%%%%%%%%%%%%%%%%%%%%%%%%%%%%%%%%%%%%%%%%%%%%%%
   & & \\
    & $ 3({\bf 1},{\overline {\bf 15}}) ( -1,-{1},+{1})_L$ &
     $3 ({\bf 1},{\bf 1},{\bf 15})(0,0,+2)_L$ \\
 $T3$ &       $3({\bf 1},{\bf 6})(+2,+1,+1)_L$ &
      $ 3({\bf 1},{\bf 1}, {\bf 6})(\pm 1,\pm {1},+1)_L$ \\
    &  $3 ({\bf 1},{\overline {\bf 6}})(+1,0,+2)_L$ & \\ 
  & & \\
   \hline
%%%%%%%%%%%%%%%%%%%%%%%%%%%%%%%%%%%%%%%%%%%%%%%%%%%%%%%%%%%%%%%%%%%%%%%%%%%%%%%
 & & \\ 
  $T6$ &       $6 ({\bf 1},{\overline {\bf 6}})(-2,+1,-{1})_L$ & 
     $6 ({\bf 1},{\bf 1},{\overline {\bf 6}})(\pm 1,\mp {1},-1)_L$ \\
  & $6 ({\bf 1},{\bf 15})(+1,-1,-{1})_L$ & \\
  & & \\
   \hline 
%%%%%%%%%%%%%%%%%%%%%%%%%%%%%%%%%%%%%%%%%%%%%%%%%%%%%%%%%%%%%%%%%%%%%%%%%%%%%%%
 & & \\   
     $T2$   & $2 ({\bf 1},{\bf 1})(-3, {\pm 3},{\pm 3})_L$ &
     $2 ({\bf 1},{\bf 2},{\bf 1})({\pm 1},{\mp 3},+3)_L$ \\
 & $2 ({\overline {\bf 3}},{\bf 1})(+3,-3,+1)_L$ & \\
 & & \\
 \hline
%%%%%%%%%%%%%%%%%%%%%%%%%%%%%%%%%%%%%%%%%%%%%%%%%%%%%%%%%%%%%%%%%%%%%%%%%%%%%%%
 $U(1)$  &
    $({1\over{3\sqrt{2}}},~{1\over{2\sqrt{3}}},~{1\over 
    {2\sqrt{3}}})$ &
  $({1\over 2},~{1\over {2\sqrt{3}}},~{1\over {3\sqrt{2}}})$~ \\
%%%%%%%%%%%%%%%%%%%%%%%%%%%%%%%%%%%%%%%%%%%%%%%%%%%%%%%%%%%%%%%%%%%%%%%%%%%%%%%
\end{tabular}
%%%%%%%%%%%%%%%%%%%%%%%%%%%%%%%%%%%%%%%%%%%%%%%%%%%%%%%%%%%%%%%%%%%%%%%%%%%%%%%
\caption{The massless spectra of the two $SU(5)$ models $S3$ and 
$S4$ with gauge symmetries 
$SU(3)_1 \otimes SU(6)_3 \otimes U(1)^3$ and 
$SU(2)_1 \otimes SU(2)_1 \otimes SU(6)_3 \otimes U(1)^3$, respectively. 
Note that double signs (as in $({\bf 1},{\bf 2},{\bf 1})(\pm 1,\pm 3, -3)_L$)
are correlated. The $U(1)$ normalization 
radii are given at the bottom of the table.
The graviton, dilaton and gauge supermultiplets are not shown.}

\end{table}
%%%%%%%%%%%%%%%%%END OF TABLE III%%%%%%%%%%%%%%%%%%%
%%%%%%%%%%%%%%%%%%%%%%%%%%%%%%%%%%%%%%%%%%%%%%%%%%%%%%%%%%%%%%

%%%%%%%%%%%%%%%%%% TABLE I V%%%%%%%%%%%%%%%%%%%%%%
%%%%%%%%%%%%%%%%%%%%%%%%%%%%%%%%%%%%%%%%%%%%%%%%%%%%%%%%%%%%%%%%%%%%%%%%%%%%%%%
\begin{table}[t]
\begin{tabular}{|c|l|l|}
%%%%%%%%%%%%%%%%%%%%%%%%%%%%%%%%%%%%%%%%%%%%%%%%%%%%%%%%%%%%%%%%%%%%%%%%%%%%%%%
  &  $S1(1,1)$ & $F11(1,0)$ \\
 M    &$SU(2) \otimes SU(6) \otimes U(1)^4$   
 &  $SU(2)^2 \otimes SU(5) \otimes U(1)^4$ \\ \hline
%%%%%%%%%%%%%%%%%%%%%%%%%%%%%%%%%%%%%%%%%%%%%%%%%%%%%%%%%%%%%%%%%%%%%%%%%%%%%%%
  & &\\
      & $({\bf 1},{\bf 35})(0,0,0,0)_L$& $ ({\bf 1},{\bf 1},{\bf 24})(0,0,0,0)_L$ \\
  $U$ & $2 ({\bf 1},{\bf 1},{\bf 1})(0,-{3},\pm 3,0)_L$
    & $2 ({\bf 1},{\bf 1},{\bf 1})(0,-{3},\pm 3,0)_L$ \\
     & $ ({\bf 1},{\bf 1},{\bf 1})(0,+6,0,0)_L$ & $ ({\bf 1},{\bf 1},{\bf 1})(0,+6,0,0)_L$ \\
    & $ ({\bf 1},{\bf 1},{\bf 1})(0,0,0,0)_L$ & $ ({\bf 1},{\bf 1},{\bf 1})(0,0,0,0)_L$ \\
     & & $ ({\bf 1},{\bf 3},{\bf 1})(0,0,0,0)_L$ \\
 & &\\
 \hline
%%%%%%%%%%%%%%%%%%%%%%%%%%%%%%%%%%%%%%%%%%%%%%%%%%%%%%%%%%%%%%%%%%%%%%%%%%%%%%%
%%%%%%%%%%%%%%%%%%%%%%%%%%%%%%%%%%%%%%%%%%%%%%%%%%%%%%%%%%%%%%%%%%%%%%%%%%%%%%%
  & &\\
                 & $ ({\bf 1},{\bf 15})(0,-{2},0 ,0)_L$    
                 & $ ({\bf 1},{\bf 2},{\overline {\bf 5}})(0,-{2},0 ,+1)_L$  \\ %%%%%%%%%
                       & $ ({\bf 1},{\overline {\bf 6}})(0,-2,0,+1)_L$     
                       & $ ({\bf 1},{\bf 2},{\bf 1})(0,-2,0,-5)_L$  \\ %%%%%%
                 & $ ({\bf 1},{\overline {\bf 6}})(0,-2,0,-1)_L$
                 & $ ({\bf 1},{\bf 1},{\bf 10})(0,-2,0,-2)_L$\\ %%%%%%%%
                        & &  $ ({\bf 1},{\bf 1},{\bf 5})(0,-2,0,+4)_L$ \\ %%%%%%%
 $T3$        
                        & $2({\bf 1},{\bf 5})(0,+1,\pm 1 ,0)_L$
                        & $2({\bf 1},{\bf 2},{\overline {\bf 5}})(0,+1,\pm 1 ,+1)_L$  \\ %%%%%%%%%
                 & $2 ({\bf 1},{\overline {\bf 6}})(0,+1, \pm 1,+1)_L$ 
                 & $2 ({\bf 1},{\bf 2},{\bf 1})(0,+1, \pm 1,-5)_L$  \\ %%%%%%
                        & $2 ({\bf 1},{\overline {\bf 6}})(0,+1, \pm 1,-1)_L$
                        & $ 2({\bf 1},{\bf 1},{\bf 10})(0,+1, \pm 1,-2)_L$\\ %%%%%%%%
                 & & $2 ({\bf 1},{\bf 1},{\bf 5})(0,+1,\pm 1,+4)_L$ \\ %%%%%%%
    & &\\
 \hline
%%%%%%%%%%%%%%%%%%%%%%%%%
   & &\\
                    & $ ({\bf 1},{\overline {\bf 15}})(\pm 1,-1,0,0)_L$
                    & $ ({\bf 1},{\bf 2},{{\bf 5}})(\pm 1,-{1},0 ,-1)_L$  \\ %%%%%%%%%
                               & $ ({\bf 1},{\bf 6})(\pm 1,-1,0,+1)_L$
                               & $ ({\bf 1},{\bf 2},{\bf 1})(\pm 1,-1,0,+5)_L$  \\ %%%%%%
                    & $ ({\bf 1},{\bf 6})(\pm 1,-1,0,-1)_L$
                    & $ ({\bf 1},{\bf 1},{\overline {\bf 10}})(\pm 1,-1,0,+2)_L$\\ %%%%%%%%
                               & & $ ({\bf 1},{\bf 1},{\overline {\bf 5}})(\pm 1,-1,0,-4)_L$ \\ %%%%%%%
  & &\\
 \hline
%%%%%%%%%%%%%%%%%%%%%%%%%%%%%%%%%%%%%%%%%%%%%%%%%%%%%%%%%%%%%%%%%%%%%%%%%%%%%%%
  & &\\
   $T2$       & $({\bf 2},{\bf 1},{\bf 1})(0,0,{\pm 3},0)_L$
   & $({\bf 2},{\bf 1},{\bf 1})(0,0,{\pm 3},0)_L$ \\
    & $({\bf 1},{\bf 1},{\bf 1})(\pm {3},+{3},0,0)_L$ 
   & $({\bf 1},{\bf 1},{\bf 1})(\pm {3},+{3},0,0)_L$ \\
 & &\\
\hline
%%%%%%%%%%%%%%%%%%%%%%%%%%%%%%%%%%%%%%%%%%%%%%%%%%%%%%%%%%%%%%%%%%%%%%%%%%%%%%%
  & &\\
 $U(1)$      & $(1/\sqrt{6}, 1/{3\sqrt{2}},1/\sqrt{6},1/\sqrt{6})$
      & $(1/\sqrt{6}, 1/{3\sqrt{2}},1/\sqrt{6},1/{3\sqrt{10}})$ \\
%%%%%%%%%%%%%%%%%%%%%%%%%%%%%%%%%%%%%%%%%%%%%%%%%%%%%%%%%%%%%%%%%%%%%%%%%%%%%%%
\end{tabular}
%%%%%%%%%%%%%%%%%%%%%%%%%%%%%%%%%%%%%%%%%%%%%%%%%%%%%%%%%%%%%%%%%%%%%%%%%%%%%%%
\caption{The massless spectra of the $S1(1,1)$ and 
$F11(1,0)$ models with gauge groups $SU(2)_1 \otimes SU(6)_3 \otimes U(1)^4$ and 
$SU(2)_1 \otimes SU(2)_3 \otimes SU(5)_3 \otimes U(1)^4$, respectively.
The $U(1)$ normalization radii are given at the bottom of the Table.
The gravity, dilaton and gauge supermultiplets are not shown.}

\end{table}
%%%%%%%%%%%%%%% END OF TABLE IV%%%%%%%%%%%%%%%%%%%%
%%%%
%%%%%%%%%%%%%%%%%%%%%%%%%%%%%%%%%%%%%%%%

%%%%%%%%%%%%%%%%%%%TABLE V%%%%%%%%%%%%%%%%%%%%%%%%
%%%%%%%%%%%%%%%%%%%%%%%%%%%%%%%%%%%%%%%%%%%%%%%%%%%%%%%%%%%%
\begin{table}[t]
\begin{tabular}{|c|l|l|} 
%%%%%%%%%%%%%%%%%%%%%%%%%%%%%%%%%%%%%%%%%%%%%%%%%%%%%%%%%%%%%%%%%%%%%%%%%%%%%%%
M  &  $F1$  &  $F2$ \\
   & $SU(3)   \otimes SU(5) \otimes U(1)^4$ &
    $SU(2)^2 \otimes SU(5) \otimes U(1)^4$ \\
   \hline
%%%%%%%%%%%%%%%%%%%%%%%%%%%%%%%%%%%%%%%%%%%%%%%%%%%%%%%%%%%%%%%%%%%%%%%%%%%%%%%
 & & \\
   &    $({\bf 1},{\bf 24})(0,0,0,0)_L$
   & $ ({\bf 1},{\bf 1},{\bf 24})(0,0,0,0)_L$ \\
   &    $2({\bf 1},{\bf 1})(0,0,0,0)_L$
   & $ 2({\bf 1},{\bf 1},{\bf 1})(0,0,0,0)_L$ \\
 $U$ &     $ ({\bf 1},{\bf 1}) (+6,0,0,0)_L$ 
   & $ ({\bf 1},{\bf 1},{\bf 1})(0,0,-6,0)_L$ \\
   &     $2 ({\bf 1},{\bf 1})(-3, {\pm 3},{\pm 3},0)_L$ &
     $2 ({\bf 1},{\bf 2},{\bf 1})({\pm 1},{\mp 3},+3,0)_L$ \\
   &      $2 ({\overline {\bf 3}},{\bf 1})(+3,-3,+1,0)_L$ & 
     $  ({\bf 2},{\bf 1}, {\bf 1})({\pm 2},0,+3,0)_L$\\
    & $({\bf 3},{\bf 1})(0,0,-4,0)_L$ & \\
 & & \\
 \hline
%%%%%%%%%%%%%%%%%%%%%%%%%%%%%%%%%%%%%%%%%%%%%%%%%%%%%%%%%%%%%%%%%%%%%%%%%%%%%%%
%%%%%%%%%%%%%%%%%%%%%%%%%%%%%%%%%%%%%%%%%%%%%%%%%%%%%%%%%%%%%%%%%%%%%%%%%%%%%%%
 & & \\
   &      $6 ({\bf 1},{\bf 10})(+1,-1,-{1},-2)_L$ &
     $3 ({\bf 1},{\bf 1},{\bf 10})(0,0,+2,-2)_L$ \\
   &      $6 ({\bf 1},{\bf 5})(+1,-1,-1,+4)_L$ &
     $3 ({\bf 1},{\bf 1},{\bf 5})(0,0,+2,+{4})_L$ \\
  $T3$ &     $6 ({\bf 1},{\bf 1})(-2,+1,-1,-{5})_L$ & 
     $6 ({\bf 1},{\bf 1},{\bf 1})(\pm 1,\mp {1},{-1},-5)_L$ \\
   &      $6 ({\bf 1},{\overline {\bf 5}})(-2,+1,-{1},+1)_L$ & 
     $6 ({\bf 1},{\bf 1},{\overline {\bf 5}})(\pm 1,\mp {1},-1,+{1})_L$ \\
   &      $3 ({\bf 1},{\bf 1})(+1,0,+2,-{5})_L$ &
      \\
   &       $3 ({\bf 1},{\overline {\bf 5}})(+1,0,+2,+1)_L$ & \\
 & & \\
      \hline
%%%%%%%%%%%%%%%%%%%%%%%%%%%%%%%%%%%%%%%%%%%%%%%%%%%%%%%%%%%%%%%%%%%%%%%%%%%%%%%
 & & \\
   &$3({\bf 1},{\bf 1})(+2,+1,+1,+5)_L$
      &
      $ 3({\bf 1},{\bf 1},{\bf 1}) (\pm 1,\pm {1},+1,+{5})_L$ \\
 $T6$  &      $3({\bf 1},{\bf 5})(+2,+1,+1,-1)_L$ &
      $ 3({\bf 1},{\bf 1}, {\bf 5})(\pm 1,\pm {1},+1,-1)_L$ \\
   &     $ 3({\bf 1},{\overline {\bf 10}}) ( -1,-{1},+{1},+2)_L$
       &
      \\
   & $ 3({\bf 1},{\overline {\bf 5}})( -1,-{1},+1, -4)_L$
    & \\
 & & \\
 \hline 
%%%%%%%%%%%%%%%%%%%%%%%%%%%%%%%%%%%%%%%%%%%%%%%%%%%%%%%%%%%%%%%%%%%%%%%%%%%%%%%
 & & \\
   &        $({\bf 3},{\bf 1})(\pm 3,-3,-1,0)_L$ &
  $({\bf 2},{\bf 2},{\bf 1})(\pm 1,\mp 3,0,0)_L$ \\
 $T2$    &    $({\overline {\bf 3}},{\bf 1})(-3,+3,+1,0)_L$ &
 $({\bf 1},{\bf 2},{\bf 1})(\pm 1,\pm 3, -3,0)_L$ \\
   &    $ ({\bf 1},{\bf 1})(+3,\pm 3,\mp 3,0)_L$ & \\
 & & \\
 \hline
%%%%%%%%%%%%%%%%%%%%%%%%%%%%%%%%%%%%%%%%%%%%%%%%%%%%%%%%%%%%%%%%%%%%%%%%%%%%%%%
 $U(1)$ &     $({1\over{3\sqrt{2}}},~{1\over{2\sqrt{3}}},~{1\over 
    {2\sqrt{3}}},~{1\over {3\sqrt{10}}})$ &
  $({1\over 2},~{1\over {2\sqrt{3}}},~{1\over {3\sqrt{2}}},~{1\over
  {3\sqrt{10}}})$~ \\
%%%%%%%%%%%%%%%%%%%%%%%%%%%%%%%%%%%%%%%%%%%%%%%%%%%%%%%%%%%%%%%%%%%%%%%%%%%%%%%
\end{tabular}
%%%%%%%%%%%%%%%%%%%%%%%%%%%%%%%%%%%%%%%%%%%%%%%%%%%%%%%%%%%%%%%%%%%%%%%%%%%%%%%
\caption{The massless spectra of the two $SU(5)$ models $F1$ and 
$F2$ with gauge symmetries 
 $SU(3)_1 \otimes SU(5)_3 \otimes U(1)^4$ and 
$SU(2)_1 \otimes SU(2)_1 \otimes SU(5)_3 \otimes U(1)^4$, respectively. 
Note that double signs (as in $({\bf 1},{\bf 2},{\bf 1})(\pm 1,\pm 3, -3,0)_L$)
are correlated. The $U(1)$ normalization 
radii are given at the bottom of the table.
The graviton, dilaton and gauge supermultiplets are not shown.}

\end{table}
%%%%%%%%%%%%%%%%%%END OF TABLE V%%%%%%%%%%%%%%%%%%
%%%%%%%%%%%%%%%%%%%%%%%%%%%%%%%%%%%%%%%%%%%%%%%%%%%%%%%%%%%%%

%%%%%%%%%%%%%%%%%%%TABLE VI%%%%%%%%%%%%%%%%%%%%%%%%
%%%%%%%%%%%%%%%%%%%%%%%%%%%%%%%%%%%%%%%%%%%%%%%%%%%%%%%%%%%%
\begin{table}[t]
\begin{tabular}{|c|l|l|} 
%%%%%%%%%%%%%%%%%%%%%%%%%%%%%%%%%%%%%%%%%%%%%%%%%%%%%%%%%%%%%%%%%%%%%%%%%%%%%%%
M  &  $F3$  &  $F4$ \\
   & $SU(3)   \otimes SU(5) \otimes U(1)^4$ &
    $SU(2)^2 \otimes SU(5) \otimes U(1)^4$ \\
   \hline
%%%%%%%%%%%%%%%%%%%%%%%%%%%%%%%%%%%%%%%%%%%%%%%%%%%%%%%%%%%%%%%%%%%%%%%%%%%%%%%
 & & \\
   &    $({\bf 1},{\bf 24})(0,0,0,0)_L$
   & $ ({\bf 1},{\bf 1},{\bf 24})(0,0,0,0)_L$ \\
   &    $2({\bf 1},{\bf 1})(0,0,0,0)_L$
   & $ 2({\bf 1},{\bf 1},{\bf 1})(0,0,0,0)_L$ \\
 $U$ &     $ ({\bf 1},{\bf 1}) (+6,0,0,0)_L$ 
   & $ ({\bf 1},{\bf 1},{\bf 1})(0,0,-6,0)_L$ \\
  &        $({\bf 3},{\bf 1})(\pm 3,-3,-1,0)_L$ &
  $({\bf 2},{\bf 2},{\bf 1})(\pm 1,\mp 3,0,0)_L$ \\
   &    $({\overline {\bf 3}},{\bf 1})(-3,+3,+1,0)_L$ &
 $({\bf 1},{\bf 2},{\bf 1})(\pm 1,\pm 3, -3,0)_L$ \\
   &   $ ({\bf 1},{\bf 1})(+3,\pm 3,\mp 3,0)_L$    & 
     $  ({\bf 2},{\bf 1}, {\bf 1})({\pm 2},0,+3,0)_L$\\
    & $({\bf 3},{\bf 1})(0,0,-4,0)_L$ & \\
 & & \\
 \hline
%%%%%%%%%%%%%%%%%%%%%%%%%%%%%%%%%%%%%%%%%%%%%%%%%%%%%%%%%%%%%%%%%%%%%%%%%%%%%%%
%%%%%%%%%%%%%%%%%%%%%%%%%%%%%%%%%%%%%%%%%%%%%%%%%%%%%%%%%%%%%%%%%%%%%%%%%%%%%%%
 & & \\
  & $3({\bf 1},{\bf 1})(+2,+1,+1,+5)_L$
      &
      $ 3({\bf 1},{\bf 1},{\bf 1}) (\pm 1,\pm {1},+1,+{5})_L$ \\
  &      $3({\bf 1},{\bf 5})(+2,+1,+1,-1)_L$ &
      $ 3({\bf 1},{\bf 1}, {\bf 5})(\pm 1,\pm {1},+1,-1)_L$ \\
   $T3$ &     $ 3({\bf 1},{\overline {\bf 10}}) ( -1,-{1},+{1},+2)_L$
       & $3 ({\bf 1},{\bf 1},{\bf 10})(0,0,+2,-2)_L$
      \\
   & $ 3({\bf 1},{\overline {\bf 5}})( -1,-{1},+1, -4)_L$
    &  $3 ({\bf 1},{\bf 1},{\bf 5})(0,0,+2,+{4})_L$\\
   &      $3 ({\bf 1},{\bf 1})(+1,0,+2,-{5})_L$ &
      \\
   &       $3 ({\bf 1},{\overline {\bf 5}})(+1,0,+2,+1)_L$ & \\
 & & \\
      \hline
%%%%%%%%%%%%%%%%%%%%%%%%%%%%%%%%%%%%%%%%%%%%%%%%%%%%%%%%%%%%%%%%%%%%%%%%%%%%%%%
 & & \\
  &      $6 ({\bf 1},{\bf 10})(+1,-1,-{1},-2)_L$ &
      \\
   &      $6 ({\bf 1},{\bf 5})(+1,-1,-1,+4)_L$ &
      \\
  $T6$ &     $6 ({\bf 1},{\bf 1})(-2,+1,-1,-{5})_L$ & 
     $6 ({\bf 1},{\bf 1},{\bf 1})(\pm 1,\mp {1},{-1},-5)_L$ \\
   &      $6 ({\bf 1},{\overline {\bf 5}})(-2,+1,-{1},+1)_L$ & 
     $6 ({\bf 1},{\bf 1},{\overline {\bf 5}})(\pm 1,\mp {1},-1,+{1})_L$ \\
  & & \\
 \hline 
%%%%%%%%%%%%%%%%%%%%%%%%%%%%%%%%%%%%%%%%%%%%%%%%%%%%%%%%%%%%%%%%%%%%%%%%%%%%%%%
 & & \\
   $T2$ &     $2 ({\bf 1},{\bf 1})(-3, {\pm 3},{\pm 3},0)_L$ &
     $2 ({\bf 1},{\bf 2},{\bf 1})({\pm 1},{\mp 3},+3,0)_L$ \\
    &   $2 ({\overline {\bf 3}},{\bf 1})(+3,-3,+1,0)_L$  & \\
 & & \\
 \hline
%%%%%%%%%%%%%%%%%%%%%%%%%%%%%%%%%%%%%%%%%%%%%%%%%%%%%%%%%%%%%%%%%%%%%%%%%%%%%%%
 $U(1)$ &     $({1\over{3\sqrt{2}}},~{1\over{2\sqrt{3}}},~{1\over 
    {2\sqrt{3}}},~{1\over {3\sqrt{10}}})$ &
  $({1\over 2},~{1\over {2\sqrt{3}}},~{1\over {3\sqrt{2}}},~{1\over
  {3\sqrt{10}}})$~ \\
%%%%%%%%%%%%%%%%%%%%%%%%%%%%%%%%%%%%%%%%%%%%%%%%%%%%%%%%%%%%%%%%%%%%%%%%%%%%%%%
\end{tabular}
%%%%%%%%%%%%%%%%%%%%%%%%%%%%%%%%%%%%%%%%%%%%%%%%%%%%%%%%%%%%%%%%%%%%%%%%%%%%%%%
\caption{The massless spectra of the two $SU(5)$ models $F3$ and 
$F4$ with gauge symmetries 
 $SU(3)_1 \otimes SU(5)_3 \otimes U(1)^4$ and 
$SU(2)_1 \otimes SU(2)_1 \otimes SU(5)_3 \otimes U(1)^4$, respectively. 
Note that double signs (as in $({\bf 1},{\bf 2},{\bf 1})(\pm 1,\pm 3, -3,0)_L$)
are correlated. The $U(1)$ normalization 
radii are given at the bottom of the table.
The graviton, dilaton and gauge supermultiplets are not shown.}

\end{table}
%%%%%%%%%%%%%%%%%%END OF TABLE VI%%%%%%%%%%%%%%%%%%
%%%%%%%%%%%%%%%%%%%%%%%%%%%%%%%%%%%%%%%%%%%%%%%%%%%%%%%%%%%%%

%%%%%%%%%%% Table VII %%%%%%%%%%%%%%%%%%%%%%%%%%%%%
%%%%%%%%%%%%%%%%%%%%%%%%%%%%%%%%%%%%%%%%%%%%%%%%%%%%%%%%%%%%%%%%%%%%%%%%%%%%%%%
\begin{table}[t]
\begin{tabular}{|c||l|l|} 
%%%%%%%%%%%%%%%%%%%%%%%%%%%%%%%%%%%%%%%%%%%%%%%%    
  & $F5$, $F7$ & $F6$, $F8$ \\
  & $SU(3)_1 \otimes SU(5)_3 \otimes U(1)^4$ & $SU(2)_1 \otimes SU(2)_1 \otimes SU(5)_3  
  \otimes U(1)^4$\\
\hline
%%%%%%%%%%%%%%%%%%%%%%%%%%%%%%%%%%%%%%%%%%%%%%%
   & &\\
    & $4({\bf 3},{\bf 1})(0,0,+2,0)_L$ & $4({\bf 2},{\bf 1},{\bf 1})(0,0,\pm 3,0)_L$  \\
   $T2$ & $4({\overline {\bf 3}},{\bf 1})(0,0,-2,0)_L$ & $4({\bf 1},{\bf 1},{\bf 1})(\pm 2,0,0,0)_L$\\
    &  & \\ \hline
%%%%%%%%%%%%%%%%%%%%%%%%%%%%%%%%%%%%%%%%%%%%%%%%%%%%%%%%%%%%%%%%%%%%%%%%%%%%%%%
 & &\\
 $U(1)$  & $(1/ 3\sqrt{2},~1/{2\sqrt{3}},~1/{2\sqrt{3}},~1/3\sqrt{10})$ &
 $(1/{2},~1/{2\sqrt{3}},~1/{3\sqrt{2}},~1/3\sqrt{10})$\\
%%%%%%%%%%%%%%%%%%%%%%%%%%%%%%%%%%%%%%%%%%%%%%%%%%%%%%%%%%%%%%%%%%%%%%%%%%%%%%%
\end{tabular}
%%%%%%%%%%%%%%%%%%%%%%%%%%%%%%%%%%%%%%%%%%%%%%%%%%%%%%%%%%%%%%%%%%%%%%%%%%%%%%%
\caption{The states to be added to the massless spectra of the 
models $F1$ and $F3$ to obtain the massless spectra of the models $F5$ and $F7$ (first column), and to the $F2$ and $F4$ to obtain the massless spectra of the models  $F6$ and $F8$ (second column), respectively. The corresponding gauge groups are given at the top of the Table. The $U(1)$ normalization radii are given at the bottom of the Table. The additional states shown in the Table appear in the $T2$ sector.}
\end{table}

%%%%%%%%%%%%%%%%END OF TABLE VII %%%%%%%%%%%%%%%%%%%%%%
%%%%%%%%%%%%%%%%%%%%%%%%%%%%%%%%%%%%%%%%%%%%

%%%%%%%%%%%%%%%%%% TABLE VIII %%%%%%%%%%%%%%%%%%%%%%
%%%%%%%%%%%%%%%%%%%%%%%%%%%%%%%%%%%%%%%%%%%%%%%%%%%%%%%%%%%%%%%%%%%%%%%%%%%%%%%
\begin{table}[t]
\begin{tabular}{|c|l|}
%%%%%%%%%%%%%%%%%%%%%%%%%%%%%%%%%%%%%%%%%%%%%%%%%%%%%%%%%%%%%%%%%%%%%%%%%%%%%%%
  & $F9$ \\
 M & $SU(4)\otimes SU(2) \otimes SU(5) \otimes U(1)^4$ 
       \\ \hline
%%%%%%%%%%%%%%%%%%%%%%%%%%%%%%%%%%%%%%%%%%%%%%%%%%%%%%%%%%%%%%%%%%%%%%%%%%%%%%%
  & \\
    & $ ({\bf 1},{\bf 1},{\bf 24})(0,0,0,0)_L$ \\
  $U$    & $({\bf 1},{\bf 1},{\bf 1})(0,0,0,0)_L$ \\
     & $ ({\bf 4},{\bf 2},{\bf 1})(0,0,0,+3)_L$ \\
    & $ ({\overline {\bf 4}},{\bf 2},{\bf 1})(0,0,0,-3)_L$ \\
  &\\
\hline
%%%%%%%%%%%%%%%%%%%%%%%%%%%%%%%%%%%%%%%%%%%%%%%%%%%%%%%%%%%%%%%%%%%%%%%%%%%%%%%
%%%%%%%%%%%%%%%%%%%%%%%%%%%%%%%%%%%%%%%%%%%%%%%%%%%%%%%%%%%%%%%%%%%%%%%%%%%%%%%
 &  \\
          & $ ({\bf 1},{\bf 1},{\overline {\bf 10}})(+2,-2,+2,0)_L~~~
                 ({\bf 1},{\bf 1},{\overline {\bf 5}})(+2,-2,-4,0)_L$  \\
    $T3$   & $ ({\bf 1},{\bf 1},{\overline {\bf 10}})(+2,+2,+2,0)_L~~~
                 ({\bf 1},{\bf 1},{\overline {\bf 5}})(+2,+2,-4,0)_L$  \\
           & $ ({\bf 1},{\bf 1},{\overline {\bf 10}})(-4,0,+2,0)_L~~~
                 ({\bf 1},{\bf 1},{\overline {\bf 5}})(-4,0,-4,0)_L$  \\
 & \\
 \hline
%%%%%%%%%%%%%%%%%%%%%%%%%
  &\\
    & $ ({\bf 1},{\bf 1},{\bf 10})(-2,0,-2,\pm 2)_L~~~
                 ({\bf 1},{\bf 1},{\bf 5})(-2,0,+4,\pm 2)_L$  \\
     & $ ({\bf 1},{\bf 1},{\bf 10})(+1,-1,-2,\pm 2)_L~~~
                 ({\bf 1},{\bf 1},{\bf 5})(+1,-1,+4,\pm 2)_L$  \\
     & $ ({\bf 1},{\bf 1},{\bf 10})(+1,+1,-2,\pm 2)_L~~~
                 ({\bf 1},{\bf 1},{\bf 5})(+1,+1,+4,\pm 2)_L$  \\
   $T6$   & $ ({\bf 1},{\bf 2},{\overline {\bf 5}})(-2,0,+1,0)_L~~~
                 ({\bf 1},{\bf 1},{\bf 1})(-2,0,-5,0)_L$  \\
     & $ ({\bf 1},{\bf 2},{\overline {\bf 5}})(+1,-1,+1,0)_L~~~
                 ({\bf 1},{\bf 1},{\bf 1})(+1,-1,-5,0)_L$  \\
     & $ ({\bf 1},{\bf 2},{\overline {\bf 5}})(+1,+1,+1,0)_L~~~
                 ({\bf 1},{\bf 1},{\bf 1})(+1,+1,-5,0)_L$  \\
  & \\
 \hline
%%%%%%%%%%%%%%%%%%%%%%%%%%%%%%%%%%%%%%%%%%%%%%%%%%%%%%%%%%%%%%%%%%%%%%%%%%%%%%%
 & \\
     & $ ({\bf 4},{\bf 2},{\bf 1})(0,0,0,-3)_L$ \\
    & $ ({\overline {\bf 4}},{\bf 2},{\bf 1})(0,0,0,+3)_L$ \\
     $T2$ & $ ({\bf 6},{\bf 1},{\bf 1})(+6,0,0,0)_L$ \\
     & $ ({\bf 6},{\bf 1},{\bf 1})(-3,+3,0,0)_L$ \\
     & $ ({\bf 6},{\bf 1},{\bf 1})(-3,-3,0,0)_L$ \\
  & \\
\hline
%%%%%%%%%%%%%%%%%%%%%%%%%%%%%%%%%%%%%%%%%%%%%%%%%%%%%%%%%%%%%%%%%%%%%%%%%%%%%%%
 &  \\
 $U(1)$      & $(1/6, 1/{2\sqrt{3}},1/{3\sqrt{10}},1/{2\sqrt{3}})$ \\
%%%%%%%%%%%%%%%%%%%%%%%%%%%%%%%%%%%%%%%%%%%%%%%%%%%%%%%%%%%%%%%%%%%%%%%%%%%%%%%
\end{tabular}
%%%%%%%%%%%%%%%%%%%%%%%%%%%%%%%%%%%%%%%%%%%%%%%%%%%%%%%%%%%%%%%%%%%%%%%%%%%%%%%
\caption{The massless spectrum of the $F9$ model
with gauge group $SU(4)_1 \otimes SU(2)_1 \otimes SU(5)_3 \otimes U(1)^4$.
The $U(1)$ normalization radii are given at the bottom of the Table.
The gravity, dilaton and gauge supermultiplets are not shown.}
\end{table}
%%%%%%%%%%%%%%% END OF TABLE VIII %%%%%%%%%%%%%%%%%%%
%%%%%%%%%%%%%%%%%%%%%%%%%%%%%%%%%%%%%%%%%%%%%

%%%%%%%%%%%%%%%%END OF TABLES%%%%%%%%%%%%%%%%%%%%
%%%%%%%%%%%%%%%%%%%%%%%%%%%%%%%%%%%%%%%%%%%%%

\end{document}